# Speech Emotion Recognition System by Quaternion Nonlinear Echo State Network

Fatemeh Daneshfar, Seyed Jahanshah Kabudian

Department of Computer Engineering, Razi University, Kermanshah, IRAN
F.daneshfar@uok.ac.ir, Kabudian@razi.ac.ir

Echo state network (ESN) is a powerful and efficient tool for displaying dynamic data. However, many existing ESNs have limitations for properly modeling high-dimensional data. The most important limitation of these networks is the high memory consumption due to their reservoir structure, which has prevented the increase of reservoir units and the maximum use of special capabilities of this type of networks. One way to solve this problem is to use quaternion algebra. Because quaternions have four different dimensions, high-dimensional data are easily represented and, using Hamilton's multiplication, with fewer parameters than real numbers, make external relations between the multidimensional features easier. In addition to the memory problem in the ESN network, the linear output of the ESN network poses an indescribable limit to its processing capacity, as it cannot effectively utilize higher-order statistics of features provided by the nonlinear dynamics of reservoir neurons. In this research, a new structure based on ESN is presented, in which quaternion algebra is used to compress the network data with the simple split function, and the output linear combiner is replaced by a multidimensional bilinear filter. This filter will be used for nonlinear calculations of the output layer of the ESN. In addition, the two-dimensional principal component analysis (2dPCA) technique is used to reduce the number of data transferred to the bilinear filter. In this study, the coefficients and the weights of the quaternion nonlinear ESN (QNESN) are optimized using genetic algorithm (GA). In order to prove the effectiveness of the proposed model compared to the previous methods, experiments for speech emotion recognition (SER) have been performed on EMODB, SAVEE and IEMOCAP speech emotional datasets. Comparisons show that the proposed QNESN network performs better than the ESN and most currently SER systems.

**Keywords:** Echo state network, Quaternion algebra, Bilinear filter, Speech emotion recognition

## 1. Introduction

Reservoir computing (RC), [1] is one of the most widely used and effective methods in training recursive neural networks (RNNs). These networks have a dynamic repository that can be easily used to process complex data. An RC network usually consists of three components: an input layer, a large RNN layer (called a reservoir), and a linear output layer in which the weights of the input layer are hidden and the weights of the hidden layer (return weights) are randomly assigned. They become constant throughout the learning period. Repository computing networks can avoid the complex RNN training process, which is based on gradient descent. While learning the dynamic state generated from the recursive hidden layer, performing well in identifying nonlinear systems, signal processing systems and time series prediction systems [3,4].

ESNs are the most popular RC networks due to their relative simplicity and strong theoretical reasoning [6-9] and have had many successful applications so far [10]. The reservoir in the echo state network is initialized with sparse connections and limitations in the spectral radius, which ensures long-term and rich echo sate dynamics [9,1,11]. This repository can be viewed as a nonlinear time core that can map the sequence of inputs to a high-dimensional space. In this learning space, it will be a simple linear regression operation from the repository to the outputs. Hence, ESNs are a powerful and efficient tool for dynamic display. However, many existing ESNs have limitations to the proper modeling of large data sets. The most important limitation of these networks is their high memory consumption due to their reservoir structure, which prevents the increase of reservoir units and maximum use of capabilities of this type of network. One way to solve this problem is to use quaternion algebra and Hamilton's multiplication for network data. Since quaternions have four different dimensions, using the Hamilton's multiplication property, they can easily display compact data with high dimensions and discover the relationships between features with fewer parameters (less memory consumption). Recently, limited work has been done on the use of quaternion algebra for speech recognition [12-14], and according to the author's knowledge, this is the first time that this type of complex numbers has been used in the field of speech emotion recognition. Although quaternion algebra has already been used in echo state network design [15], the model presented is based on the use of a nonlinear quaternion activation function with local analytical properties in which the nonlinear gradient descent algorithm is used to find the weights of the output layer. Since in this current paper the output weights are estimated based on a non-gradient way, there is no need for a complex and time-consuming process to produce gradients and to derive derivative conditions of the quaternion activity function. Therefore, in this current research, a simpler concept called split function [16] will be used as an activity function, which is actually a function with real value that is applied as a fraction to a quaternion number.

In addition to the memory consumption problem in the echo state network, in all ESN models presented so far, the linear output of the network creates an indescribable limitation to the ESN processing capability, as it cannot effectively use the higher order statistics of the signals provided by the repository. Therefore, one of the effective methods is to study and propose different schemes for combining the signals provided by the reservoir in order to approximate the desired input behavior and produce output more carefully. In this regard, using a nonlinear structure in the output layer can be a good choice. In this research, a multidimensional bilinear filter will be used in ESN network for nonlinear output layer calculations based on quaternion algebra. In addition, in this study, before transferring the network reservoir state to a multidimensional bilinear filter, using the two-dimensional principal component analysis (2dPCA) technique, we prevent the number of bilinear filter coefficients from increasing too much and its complexity from increasing.

The method proposed in this paper has been investigated in a SER system. In this system, after extracting short-term features, statistical data are used to find related frame features, taking into account that the extracted features are at the frame level, while the evaluation of SER systems is usually based on emotional speech. In this paper, the long-term features of mean, standard deviation, skewness, and kurtosis (which well-describe four different perspectives of a signal) are considered as different dimensions of quaternion numbers to discover both the internal relations between these statistical data and the external relations between different time frames by Hamilton's multiplication. In addition, the bilinear filter coefficients as well as the weights of the QNESN network are optimized using the training set features and the genetic algorithm. Then this model is used to emotionally classify the features of the test set. Finally, in order to prove the effectiveness of the proposed QNESN model compared to the previous methods, experiments have been performed on three well-known sets of emotional speech. Comparisons show that QNESN generally performs better than the simple ESN as well as most recently SER systems.

Considering the above considerations, the contributions made in this paper can be summarized as follows:

1) Using the features of Gabor filter bank (GBFB) [17], single frequency cepstral coefficient (SFCC) [18] along with glottal waveform, which have not been used in recognizing speech emotions. These features are suitable for analyzing speech features, improving the results of speech / non-speech classification, providing spectral-temporal contrast at all times, reducing redundancy, better recognizing speaker-related emotions, better recognizing similar emotions, and distinguishing distinct emotions.
2) Introducing a new model of quaternion nonlinear echo state network (based on split function as activity function) for the first time in this paper, which uses quaternion algebra, will have a more compact representation of data and reduce the problem of high storage memory consumption. In addition, using its simple activity function, it will provide a simpler model for discovering relationships between input data with a smaller number of parameters.
3) Introducing a new quaternion of speech emotional features as the input of the echo sate network, which incorporates four different perspectives of speech time frames and is used for the first time for emotion recognition applications.
4) Introduce a new multidimensional bilinear filter design for the first time in this paper and use it in the output layer of the ESN network to maintain the nonlinear nature of the reservoir and present it to the output, which will increase network performance with fewer neurons, reduce complexity and simplify the mathematical model.
5) Using evolutionary algorithms to optimize all parameters of the quaternion nonlinear echo state network (since the usual methods of inverse matrix, etc. cannot be applied).
6) Reduce the dimensions of quaternion features by using two-dimensional principle component analysis that will reduce computations and increase system accuracy.

This article is organized into nine sections. The second and third sections provide an overview of current and background research. Section 4 gives a general description of the designed system. Section 5 describes how to classify using a simple echo state network, a nonlinear ESN, and a quaternion nonlinear ESN. In the sixth and seventh sections, we have evaluated and analyzed the proposed model, and finally in the eighth and ninth sections, we have discussed and concluded in this regard.

## 2. Related works
### 2.1. Quaternion Networks

Due to the ability of quaternion algebra to reduce network parameters and discover more connections between input features using Hamilton's multiplication, many recent studies have been done using it in the design of neural networks, deep networks, recurrent networks, etc. [12, 14, 19- 24]. Arena et al. (1994) first introduced the quaternion neural networks (QNN) with a special algorithm for learning it effectively in the same way as real neural networks. Following this proposal, many researches have examined the main features of QNNs and focused on each of its main components such as activation functions [25], error functions [26], initialization of parameters [21] and their architectural development for better performance in the field of quaternion multidimensional data is suggested. Because quaternion numbers are overly complex numbers that contain a real component and three separate imaginary components, they can be suitable for displaying three- and four-dimensional feature vectors such as channels (R, G, B) in image processing. For this reason, in the field of image processing, many articles have been done using quaternion algebra [19, 21, 27, 28] in which the dimensions of quaternion numbers are actually properties related to the color part of the image.

In the field of speech processing, the use of quaternion algebra is very new and so far few studies have used it. In [14] a long-term quaternion-based recursive neural network for speech recognition has been proposed in which the MFCC property, along with its derivatives, form the various dimensions of a quaternion number. In [20] a quaternion neural network is designed to detect speech across multiple channels, in which each microphone is represented by one of the dimensions of the quaternion number. In addition, a deep quaternion-based neural network has been designed to understand spoken language, and in [12] a quaternion evolutionary neural network has been used for automatic speech recognition.

One of the challenges of recent research in the field of SER is to increase the complexity of the proposed models due to the large dimensions of the feature vector. The use of quaternion algebra in neural network architecture can well reduce the complexity of models by reducing the number of parameters (reducing the size of matrices by a quarter) and ultimately have better performance. Quaternion algebra has already been used in echo state network design [15]. The model presented in this paper is based on the use of nonlinear quaternion activation function with local analytical properties in which the nonlinear gradient descent algorithm is used to find the weights of the output layer. In this model, due to the use of gradient-based methods, there will be a need for a complex and time-consuming process of gradient production and establishment of derivative conditions of the quaternion activity function.

### 2.2. Speech Emotion Recognition Systems

Recently, many studies have been published in the field of speech emotion recognition systems. In the study presented by [29], two models with Gaussian radial basis function kernel (GRBF) and a linear kernel with binary tree and a combined smooth maximum regression model for emotion classification are proposed. In [30] a semi-regulatory feature selection method is used to reduce the dimension of emotional features. The long-term properties of acoustic features have been investigated by [31] using the modulation filtering method and the excitation source component related to the speech stimulus component. The model presented by [32] uses both vocal and spectral features of speech, along with a simple Bayesian classifier. In [33-34], the residual sinusoidal amplitude and normalized cepstral coefficients are also used as emotional features. Researchers have also used speech and sound quality features as emotional features [35] and in the model proposed by [36], a randomized deep belief network (RDBN) as a deep neural network (DNN) to achieve high-level features form low-level features have been used, and finally [37] has proposed new emotional features based on the energy content of the wavelet-based time frequency distribution. In [38] provides a semi supervised generative adversarial network (SSGAN) for extracting and recognizing emotion from labeled and unlabeled data, and [39] uses an adversarial network with an auto-encoder as a classifier of emotional features. In addition, [40] has provided an adversarial auto-encoder to solve the problem of lack of emotional data and lack of proper labeling.

Meanwhile, some studies have used wavelet transform to better detect speech emotions. For example, in the model presented by [41] using wavelet packet analysis, new features of speech signal have been extracted to detect speech emotions. In [42], the features extracted using the triangular filter bank and the equivalent extracted coefficient are used to detect the emotion. Research [43] has used a weighting method to find features related to different emotions and inspired by feature selection methods such as mRMR and ReliefF, and has also used deep learning to categorize features. In the architecture proposed by [44], the vector with high dimensions of emotional features including fundamental frequency, zero cross rate, MFCC, energy and harmonic noise ratio is divided into different subsections using C-means fuzzy (FCM) clustering algorithm. They are categorized using multiple random forest algorithm. In addition, research [45] has used the active feature selection method to select effective emotional features and has shown that the selection of effective features will have a good effect on the accuracy of emotion recognition systems. In the model presented by [46], the performance of two different types of classifiers in SER systems are compared with each other. In model [47], an ensemble learning model random forest algorithm is used to find the importance of different features. In this method, the weighted binary cuckoo algorithm for speech feature selection is used to select the features and also uses the decision tree classifiers, linear differential classification, random forest and SVM. In the research presented by [48], the proposed method of discriminative non-negative matrix factorization (DSNMF) has been used to reduce the dimensions of input features. In the model proposed by [49], the Hilbert-Huang-Hurst coefficient vector (HHHC) is used as one of the nonlinear features of the audio source, due to their effect on the speech production mechanism, to better display emotional states. GMM, HMM, DNN, CNN and CRNN have also been used as categories.

Most of the mentioned methods have used deep learning and generative adversarial networks in their proposed models. The main advantage of deep learning methods is the increased accuracy of the system in recognizing emotions. But the lack of emotional speech data will overfitting deep networks. Also, the large number of parameters for setting and initialization, as well as the high learning time, are other limitations of the methods. Although generative adversarial networks are highly usable and useful for low-data set problems using additive data properties, their limitations include vanishing gradient, high training time, and the problem of non-convergence and instability in all structures. Table 1 summarizes the recent methods performed in the SER domain. In addition, Figure 1 shows the recognition rate of the recent methods according to their learning system on the EMODB dataset.

Table 1. Recent methods and their recognition rates on EMODB dataset

| Ref | Proposed method | WAR | UAR |
|---|---|---|---|
| Zhao 2020, [38] | Use the semi-supervised generative adversarial network as classifier | 65.20 | 68.00 |
| Yi 2020, [97] | User the generative adversarial network and auto-encoder as classifier | 84.49 | 83.31 |
| Latif 2020 [40] | Use the adversarial auto-encoder for feature discriminative recognition | N/A | 66.70 |
| Wang 2020, [41] | Use wavelet analysis to extract features | N/A | 79.20 |
| Sugan 2020, [42] | Feature extraction using a triangular filter bank | 77.08 | N/A |
| Vieira 2020, [49] | Use Hilbert-Huang-Hurst coefficients to extract features | 81.80 | N/A |
| Haider 2020, [45] | Use the active feature selection method to select the feature | N/A | 76.90 |
| Li 2021, [67] | Use the feature weighting method based on emotional groups to select the features | 72.19 | N/A |
| Chen 2020, [44] | Use the C-means fuzzy clustering algorithm to select feature | 85.61 | N/A |
| Song 2020 [125] | Use of robust differential sparse regression to select differential features | N/A | 86.19 |
| Zhang 2021 [47] | Using ensemble learning model random forest algorithm and weighted binary cuckoo algorithm to select superior features | 83.70 | N/A |
| Hou 2020 [48] | Use of discriminative non-negative matrix factorization for feature dimension reduction | 82.80 | 83.30 |

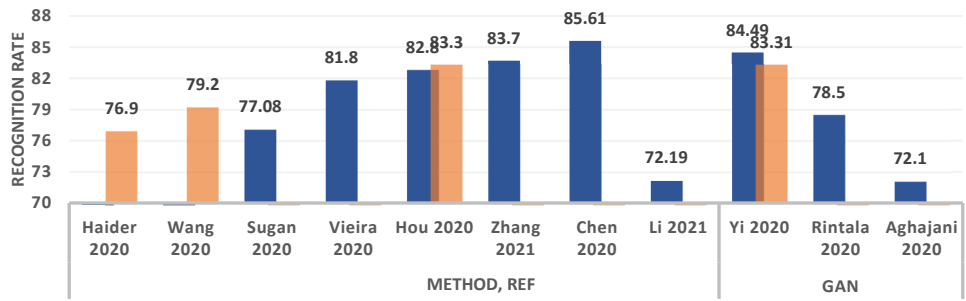

Figure 1. Recognition rates of recent methods on EMODB dataset

## 3. Background
### 3.1. Quaternion Algebra

In mathematics, quaternions are a numbering device for expanding complex numbers. They were first introduced by the Irish mathematician William Hamilton in 1843 and applied to mechanics in three-dimensional space. The set of quaternions H is equivalent to $R^4$, a four-dimensional vector space on real numbers. Thus, each element H can be uniquely written as a linear combination of these basic elements;

$$a + b\mathbf{i} + c\mathbf{j} + d\mathbf{k} \tag{1}$$

Where a, b, c and d are real numbers. The basic element 1 is the identical element H. In fact, each quaternion number, unlike real numbers, has four different dimensions, so it is easy to package multidimensional input features. This will reduce the quaternion grid parameters by a quarter. In this network, each unit will contain four times the unit of information in real networks. H has three addition operations, scalar multiplication and quaternion multiplication. The sum of two elements H is defined as the sum of them as elements $R^4$. Similarly, the product of an element H in a real number, like the product of a scalar, is defined in $R^4$.

The quaternion multiplication (Hamilton multiplication) is indicated by the symbol ⊗ and for the two quaternions Q1 and Q2 will be as follows:

$$Q_1 = [a_1 \ b_1 \ c_1 \ d_1]$$
$$Q_2 = [a_2 \ b_2 \ c_2 \ d_2] \tag{2}$$

$$Q_1 \otimes Q_2 = \begin{bmatrix} a_1 a_2 - b_1 b_2 - c_1 c_2 - d_1 d_2 \\ (a_1 b_2 + b_1 a_2 + c_1 d_2 - d_1 c_2)\mathbf{i} \\ (a_1 c_2 - b_1 d_2 + c_1 a_2 + d_1 b_2)\mathbf{j} \\ (a_1 d_2 + b_1 c_2 - c_1 b_2 + d_1 a_2)\mathbf{k} \end{bmatrix} \quad (3)$$

One of the characteristics of quaternions is that the multiplication of two quaternions has no displacement properties. But it has the property of participation and distributability. The Hamilton multiplication, due to its distributive nature, makes it possible to discover the hidden internal relationships between the components of a quaternion. In addition, it helps to encode the interdependence between input features with fewer parameters, so the use of quaternion algebra in high-dimensional space will be very cost-effective.

### 3.2. Cost Function in Quaternion Algebra

In [16] there is a concept called split function, which is actually a function with real value that will be applied to a quaternion number separately,

$$f(Q) = f(a) + f(b)\mathbf{i} + f(c)\mathbf{j} + f(d)\mathbf{k} \quad (4)$$

Arena, 1993 proved that the theorem of global approximation when using the split function in quaternion space is as true as the real number space [16]. Also in [26] a function called QMSE or mean square error in quaternion space is presented as follows and based on the split function introduced in [16].

$$E = \tfrac{1}{2} \sum_{n=1}^{N} [(t_{0n} - Y_{0n})^2 + (t_{1n} - Y_{1n})^2 + (t_{2n} - Y_{2n})^2 + (t_{3n} - Y_{3n})^2] \quad (5)$$

Where Y is the output vector of the quaternion and t is the target quaternion class of the destination. Most quaternion space classification articles [22, 26, 50] have used the QMSE function to find the best Y quaternion vector class. In this research, the proposed method in [50] has been used to find the output class.

### 3.3. Echo State Network

The echo state network is a recursive neural network with three layers: an input layer, a large recursive layer (reservoir) with fixed sparse hidden-to-hidden connections and an outer layer [7]. The overall structure of the ESN network is shown in Figure 2. In this study, we name the number of input dimensions with $N_U$, we also assume that the repository has $N_{Ru}$ nerve cell (return unit). In addition, $\vec{U}(t)$ represents the input at time t (according to the following relation),

$$U = \begin{pmatrix} u_1(1) & \cdots & u_1(T) \\ \vdots & \ddots & \vdots \\ u_{N_U}(1) & \cdots & u_{N_U}(T) \end{pmatrix}$$
$$\vec{U}(t) = [u_1(t) \ldots u_{N_U}(t)]^T \quad (6)$$

Where T is the number of input samples. Also $\vec{X}(t)$ is the state of the reservoir at time t. According to the following relation,

$$X = \begin{pmatrix} x_1(1) & \cdots & x_1(T) \\ \vdots & \ddots & \vdots \\ x_{N_{Ru}}(1) & \cdots & x_{N_{Ru}}(T) \end{pmatrix} \quad (7)$$

$$\vec{X}(t) = [x_1(t) \ ... \ x_{N_{Ru}}(t)]^T$$

Therefore, for a simple ESN, the relations would be as follows [51].

$$\vec{X}(t) = F(\vec{U}(t), \vec{X}(t-1)) = (1-\alpha)\vec{X}(t-1) + \alpha f(W_{in}\vec{U}(t) + W\vec{X}(t-1)) \quad (8)$$

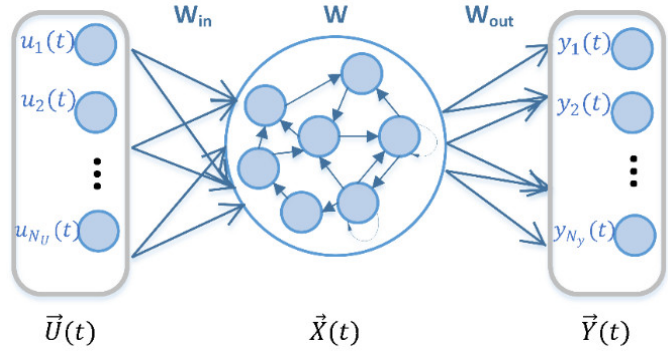

Figure 2. Structure of a simple echo state network

In relation (8), $\alpha \in [0,1]$ leaking rate and f are the activation function for reservoir units (typically, the non-linear tanh function is used). In the standard RC framework, the reservoir parameters, i.e. the input weights $W_{in}$ and the reservoir weights W, are obtained according to relations (9) and (10). These coefficients are left without training after initialization and under the stable conditions presented in [51] by analyzing the eco state specifications for different reservoir.

$$W_{in} = \begin{pmatrix} w_{in,1}(1) & \cdots & w_{in,1}(N_U) \\ \vdots & \ddots & \vdots \\ w_{in,N_{Ru}}(1) & \cdots & w_{in,N_{Ru}}(N_U) \end{pmatrix} \quad (9)$$

$$W = \begin{pmatrix} w_1(1) & \cdots & w_1(N_{Ru}) \\ \vdots & \ddots & \vdots \\ w_{N_{Ru}}(1) & \cdots & w_{N_{Ru}}(N_{Ru}) \end{pmatrix} \quad (10)$$

In relation to the output calculation, at each time step t, the state of the reservoir feeds the output layer (Figure 2). If we denote the size of the output space by $N_y$, the output in the time step t is calculated by the following linear function:

$$\vec{Y}(t) = W_{out}\vec{X}(t) \quad (11)$$

As $W_{out}$ is a readout weight matrix that is adjusted on the training set, and by methods such as inverse matrix or linear regression.

## 3.4. Nonlinear Echo State Network

Due to the limited efficiency of ESN network with linear output, one of the effective methods is to study and propose different schemes to combine the signals provided by the reservoir, in order to approximate the input behavior and produce the desired signal more accurately for the output layer. In this context, the use of a nonlinear structure in the output layer can be a good choice (Figure 3).

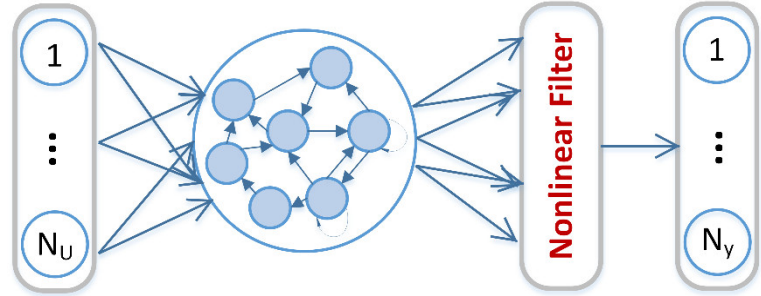

Figure 3. Nonlinear echo state network

This has been used by [52-53] in which, the Volterra filter, is used as a nonlinear filter in the output layer. As shown in relation (12), the Volterra series is a suitable method for approximating any function using N of its previous example and nonlinearly. The higher the value of N, the closer this approximation will be to reality. N actually represents the amount of memory or delay of the filter, and the order of the filter, which is $M_0$, is the number of sentences that have been multiplied consecutively, and h is an $M_0$ dimensional function, which is the Volterra filter kernel.

$$y(j) = \sum_{l_1=0}^{N-1} h_1(l_1)x(j-l_1)$$
$$+ \sum_{l_1=0}^{N-1}\sum_{l_2=0}^{N-1} h_2(l_1,l_2)x(j-l_1)x(j-l_2) + \sum_{l_1=0}^{N-1}\sum_{l_2=0}^{N-1}\sum_{l_3=0}^{N-1} h_3(l_1,l_2,l_3)x(j-l_1)x(j-l_2)x(j-l_3) + \cdots,$$

$j = 0,1,\ldots,N-1$ (12)

The most important feature of this filter is that the mapping of a Euclidean space to a scalar space, as well as the output at any time, is dependent on all previous inputs and makes it possible to use higher statistics. But one of the most important problems is that finding its kernels is complicated and time consuming and requires a lot of calculations. In addition, to estimate the nonlinear function of the Volterra filter requires the estimation of more kernels, which doubles its complexity. In 2005, to reduce the computational complexity of the Volterra filter, an adaptive bilinear filter was introduced by [53] (relation 13), in which output y(t) using the previous N-1 output, and previous input x(t), with the product of both will be achieved.

$$y(t) = \sum_{i=1}^{N-1} c_i y(t-i) + \sum_{i=0}^{N-1}\sum_{j=1}^{N-1} b_{i,j} y(t-j)x(t-i) + \sum_{i=0}^{N-1} a_i x(t-i) \qquad (13)$$

This filter, unlike Volterra, which uses only the feedforward coefficients, will use both the feedforward coefficients (matrix a) and the feedback coefficients (matrix b). In [53] it is proved that all nonlinear systems can be estimated with a good approximation using a bilinear filter and with a lower order than the Volterra filter. Figure (4) shows a one-dimensional bilinear filter in which, according to the relation shown, the N is order of the filter and its amount of memory, and the coefficients a, b and c are usually obtained using the least squares error method. The use of bilinear filter in the output layer of the ESN network will not only avoid the problems of designing the Volterra filter, but will also be able to effectively use the higher order statistics of the features provided by the reservoir. To the best of our knowledge, this has never been suggested by anyone.

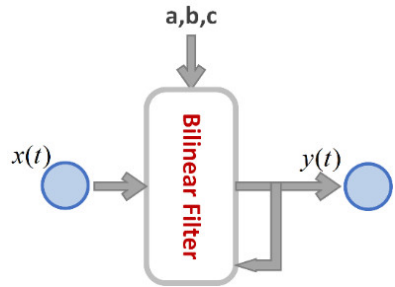

Figure 4. One-dimensional bilinear filter

So far, various bilinear filters with different applications have been proposed. Many of these models are suitable for one-input-one-output dimensions [54,55], while the SER domain is a multi-dimensional input-output space. Previously, different models based on the definition of recurrent neural network with bilinear filter in a multi-input-multi-output (MIMO) system have been proposed and used in many systems [56-57]. In this paper, for the first time, the design of a bilinear filter with multidimensional input-output has been done.

## 4. Overall System Description

As shown in Figure (5), the SER system proposed using the proposed QNESN model consists of two main sections: the preprocessing and feature extraction section, and the feature classification section. In the preprocessing phase, silent intervals are detected and eliminated. Then, the Cepstral coefficient of perceptual linear prediction (PLPC), the SFCC [18], and their first- and second-order derivatives, along with the GBFB [17], are extracted from both speech signals and glottal waveforms. They are also suitable for analyzing speech characteristics, improving the results of speech / non-speech classification, providing Spectro-temporal contrast at all times, reducing redundancy, better recognizing speaker-related emotions, better recognizing similar emotions, and recognizing distinct emotions [17]. Because glottal waveform features are significantly affected by a person's emotional state and speech style [58], in this study, glottal waveform features were used to achieve a more accurate emotional classification. Then, using the training set features and genetic algorithm, the coefficients of the proposed QNESN network are obtained, and finally, this network estimates the emotional class of the experimental speech signal. In the following, the details of each of these sections and how the QNESN model is designed will be described.

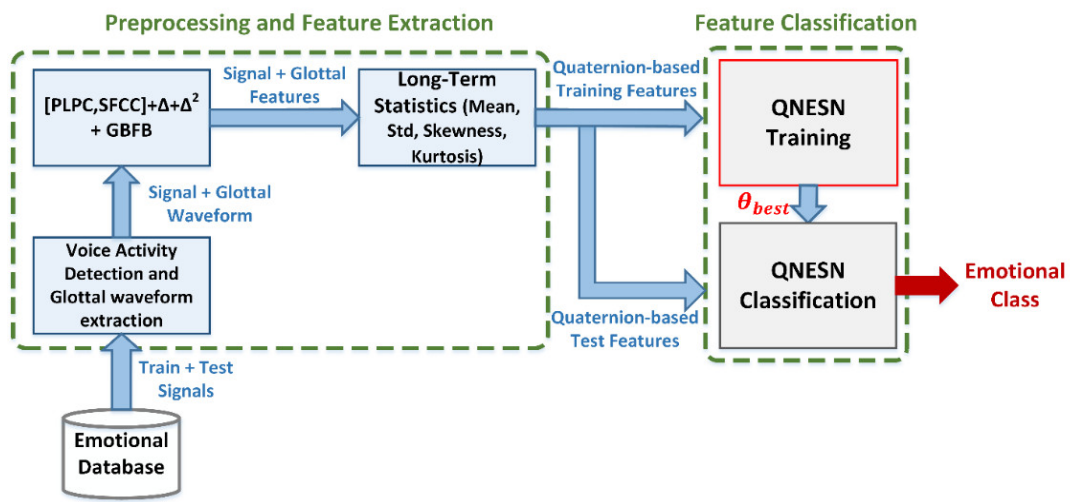

Figure 5. Overall description of the speech emotion recognition system using QNESN

## 5. Emotional Features Classification

This section explains step-by-step how to design a QNESN classifier. Initially, for better ESN network performance, we created a new connection from input to output (Figure 6). Therefore, relation (11) will change as follows,

$$\vec{Y}(t) = W_{out}(\vec{X}(t), \vec{U}(t)) \tag{14}$$

And the network output and output weight matrices are also defined as follows,

$$Y = \begin{pmatrix} y_1(1) & \cdots & y_1(T) \\ \vdots & \ddots & \vdots \\ y_{N_y}(1) & \cdots & y_{N_y}(T) \end{pmatrix} \tag{15}$$

$$\vec{Y}(t) = [y_1(1) \ \ldots \ y_{N_y}(1)]^T$$

$$W_{out} = \begin{pmatrix} w_{out,1}(1) & \cdots & w_{out,1}(N_{Ru} + N_U) \\ \vdots & \ddots & \vdots \\ w_{out,N_y}(1) & \cdots & w_{out,N_y}(N_{Ru} + N_U) \end{pmatrix} \tag{16}$$

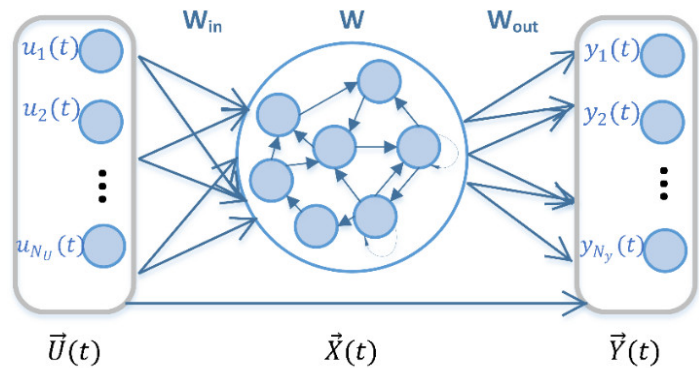

Figure 6. ESN network with new relations

Table 2 shows a comparison between network performance with the previous simple structure and the new structure in the SAVEE dataset emotion recognition rate with the features extracted in the previous section, along with a reservoir layer of 100 units. According to the results, the ESN network with the new structure has performed better than the previous structure.

Table 2. Compare ESN network performance with new and old design

| ESN Structure | WAR |
| --- | --- |
| Structure of Figure 2 | 44.37% |
| Structure of Figure 6 | 57.88% |

## 5.1. Emotional Data Preparation

Because the echo state network is highly efficient as a time series, the features extracted at the frame level will be used in this experiment. The difference is that to reduce the complexity caused by the high number of frames associated with each utterance, another framing (with a larger window length) will be performed to reduce the number of frames associated with each speech utterance.

If we display the short-term features extracted with the 10ms shift window in the feature extraction section, with a matrix V,

$$V = \begin{pmatrix} v_1(1) & \cdots & v_1(T) \\ \vdots & \ddots & \vdots \\ v_{N_V}(1) & \cdots & v_{N_V}(T) \end{pmatrix} \quad (17)$$

$$\vec{V}(t) = [v_1(t)...v_{N_V}(t)]^T$$

Then, by applying the statistical functions (medium-term features) of the mean, standard deviation, skewness and kurtosis with 2W frame window length and K frame shift window, on the V vector, the final input vector will be obtained according to the following relation:

$$\vec{U}(t) = [\mathrm{mean}(\vec{V}(t-W:t+W)) \quad \mathrm{std}(\vec{V}(t-W:t+W)) \\ \mathrm{skewness}(\vec{V}(t-W:t+W)) \quad \mathrm{kurtosis}(\vec{V}(t-W:t+W))]^T_{1 \times N_U} \quad (18)$$

## 5.2. Echo State Network Hyperparameters

Important hyper-parameters used to initialize an ESN network are:

- Input Scaling (IS) - The input scale is used to generate the initial random values of the $W_{in}$ input matrix, so that the random values of its elements must have a uniform distribution between -IS and IS.
- Spectral Radius (SR) - The spectral radius is used to generate the initial random values of the reservoir weight matrix (W) using the following relation:

$$W = SR . \frac{W}{\lambda_{\max}(W)} \quad (19)$$

In this relation, the matrix W has random elements generated in the interval [-1 1], where $\lambda_{max}(W)$ represents the largest eigenvalue of that matrix. In addition, to ensure that the echo state is maintained, the spectral radius takes values less than one.

- Leaking Rate (LR) - Determines the velocity of the reservoir in response to the input.
- Regularization coefficient - (C) According to the theory [59] of the classification problem in relation (11), using the regularization coefficient C in the regularization formula introduced by [59] will improve the generalization efficiency of the ESN network.

## 5.3. Nonlinear Echo State Network using Multi-dimensional Bilinear Filter

Previously, bilinear filter relationships in one-dimensional input-output environments have been introduced in relation (13). Since the input-output of SER systems is multidimensional, to use a two-line filter, its relationships must be defined in multidimensional. The model presented by the simple ESN network using nonlinear output will be in Figure (7). In the proposed model, in the output layer, the reservoir state will be given to a multidimensional bilinear filter to obtain the final output using its coefficients. In this way, the nonlinear properties of the reservoir will be preserved and used to produce the best possible output. In this case, all the relations of the echo network will be the same as before (relations 6 to 11), except that the output will be based on relation (20) instead of relation (11). In this multidimensional model, each of the output dimensions y(t) will be obtained from the sum of N-1 of the previous output sample at that dimension, along with the reservoir state signal x(t), and the product of the two.

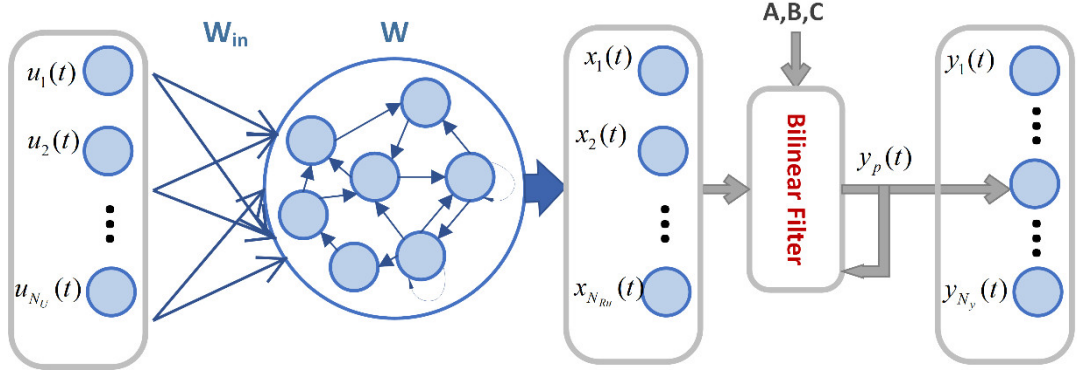

Figure 7. Nonlinear echo state network with multi-dimensional bilinear filter

$$y_p(t) = \sum_{i=1}^{N-1} c_{p,i} y_p(t-i) + \sum_{i=1}^{N-1}\sum_{j=1}^{N_{Ru}} b_{i,j}^p y_p(t-i)x_j(t) + \sum_{i=1}^{N_{Ru}} a_{p,i} x_i(t) \qquad (20)$$

$1 \leq p \leq N_y$, $N$ : filterorder

This relation will be vectorized as follows,

$$y_p(t) = \vec{C_p}\vec{Y_P^T}(t-1) + \vec{Y_P}(t-1)B_p\vec{X}(t) + \vec{A_p}\vec{X}(t) \qquad (21)$$

The filter input matrix is obtained from relation (7), where $x_i(t)$ is the i input (echo state) of sample t. Also, the output of the filter according to relation (15) is defined as vector as follows,

$$\vec{Y_p}(t-1) = [y_p(t-1) \ ... \ y_p(t-(N-1))] \qquad (22)$$

Where T is the number of input samples and $N_y$ is the number of output dimensions (y dimensions). Finally, the coefficients of the bilinear filter will be as follows (feedforward coefficient for all $N_y$ output dimensions and $N_{Ru}$ reservoir units),

$$A = \begin{pmatrix} a_{1,1} & \cdots & a_{1,N_{Ru}} \\ \vdots & \ddots & \vdots \\ a_{N_y,1} & \cdots & a_{N_y,N_{Ru}} \end{pmatrix} \qquad (23)$$

$$\vec{A_p} = [a_{p,1} \ ... \ a_{p,N_{Ru}}]$$

Cross product coefficients (for output dimensions of $p$, filter order $N$ and number of reservoir units $N_{Ru}$),

$$B^p = \begin{pmatrix} b^p_{1,1} & \cdots & b^p_{1,N_{Ru}} \\ \vdots & \ddots & \vdots \\ b^p_{(N-1),1} & \cdots & b^p_{(N-1),N_{Ru}} \end{pmatrix} \quad (24)$$

$$B = [B^1, ..., B^{N_y}]$$

Feedback coefficient,

$$C = \begin{pmatrix} c_{1,1} & \cdots & c_{1,(N-1)} \\ \vdots & \ddots & \vdots \\ c_{N_y,1} & \cdots & c_{N_y,(N-1)} \end{pmatrix} \quad (25)$$

$$\vec{C}_p = [c_{p,1} \quad \cdots \quad c_{p,(N-1)}]$$

Thus relation (21) can be written as follows,

$$y_p(t) = \vec{C}_p \vec{Y}_P^T(t-1) + B_p \vec{V}_p^T(t) + \vec{A}_p \vec{X}(t) \quad (26)$$

Which,

$$\vec{V}_p(t) = [y_p(t-1)x_1(t), ..., y_p(t-1)x_{N_{Ru}}(t), \\ y_p(t-2)x_1(t), ..., y_p(t-2)x_{N_{Ru}}(t), ..., \\ y_p(t-(N-1))x_1(t), ..., y_p(t-(N-1))x_{N_{Ru}}(t)] \quad (27)$$

And if we define the vectors W and U as follows,

$$W_p = [\vec{C}_p \quad B_p \quad \vec{A}_p] \quad (28)$$

$$\vec{U}_p(t) = [\vec{Y}_p(t-1) \quad \vec{V}_p(t) \quad \vec{X}^T(t)]^T \quad (29)$$

Then relation (26) will change as follows,

$$y_p(t) = W_p \vec{U}_p(t) \quad (30)$$

So far, various algorithms have been used to solve relation (30) and obtain filter coefficients ($W_p$ matrix).

An example of a simple multidimensional bilinear is shown in Figure (8). In this bilinear filter, the input is two-dimensional ($N_{Ru} = 2$) and the output is two-dimensional ($p = 2$) and the order of the filter is two-dimensional ($N = 2$). So, the output will be as follows,

$$y_1(t) = c_{1,1} y_1(t-1) + b^1_{1,1} y_1(t-1) x_1(t) + b^1_{1,2} y_1(t-1) x_2(t) + a_{1,1} x_1(t) + a_{1,2} x_2(t)$$

$$y_2(t) = c_{2,1} y_2(t-1) + b^2_{1,1} y_2(t-1) x_1(t) + b^2_{1,2} y_2(t-1) x_2(t) + a_{2,1} x_1(t) + a_{2,2} x_2(t)$$

And the coefficients are as,

$$A = \begin{pmatrix} a_{1,1} & a_{1,2} \\ a_{2,1} & a_{2,2} \end{pmatrix}, \begin{matrix} B^1 = [b^1_{1,1}, ..., b^1_{1,2}] \\ B^2 = [b^2_{1,1}, ..., b^2_{1,2}] \end{matrix}, \quad C = [c_{1,1}, c_{2,1}]^T$$

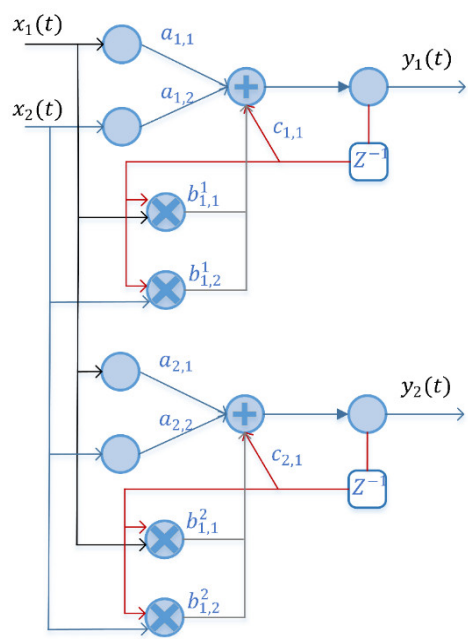

Figure 8. The structure of a multidimensional bilinear filter

## 5.4. Quaternion Nonlinear Echo State Network

One of the major problems in the design of SER systems is their high memory consumption due to the extraction of features with high emotional dimensions. Since the ESN is recursive, the weight matrix of the reservoir layer will increase in size as the reservoir units increase. The larger the number of reservoir units, the more memory is required. In addition, the larger the input size of the ESN, the larger the number of reservoir units must be to better model it. One of the available solutions to this problem is the use of quaternion algebra or high-dimensional algebra. In this study, the mean-term features of mean, standard deviation, skewness and kurtosis will be considered as different dimensions of quaternion numbers. If quaternion algebra is used in the design of a nonlinear eco-state network to detect emotion, in this architecture all input, output, coefficients and weights matrices will be quaternion (Figure (9)).

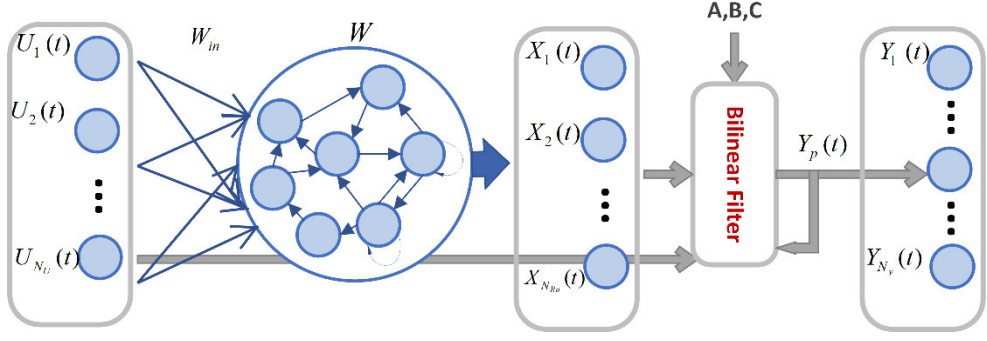

Figure 9. Quaternion nonlinear ESN

In this case, the matrix of input features is as follows, each dimension of which will include one of the mid-term properties.

$$U = \begin{pmatrix} U_1(1) & \cdots & U_1(T) \\ \vdots & \ddots & \vdots \\ U_{N_U}(1) & \cdots & U_{N_U}(T) \end{pmatrix} \tag{31}$$

$U_i(t) = u_i^1(t) + u_i^2(t)\mathbf{i} + u_i^3(t)\mathbf{j} + u_i^4(t)\mathbf{k}, \ 1 \leq i \leq N_U, \ 1 \leq t \leq T$

$$\vec{U}(t) = [\text{mean}(\vec{V}(t-W:t+W)) \quad \text{std}(\vec{V}(t-W:t+W))\mathbf{i} \\ \text{skewness}(\vec{V}(t-W:t+W))\mathbf{j} \quad \text{kurtosis}(\vec{V}(t-W:t+W))\mathbf{k}]^T \tag{32}$$

In addition, the reservoir state matrix will be defined as follows in the quaternion state.

$$X = \begin{pmatrix} X_1(1) & \cdots & X_1(T) \\ \vdots & \ddots & \vdots \\ X_{N_{Ru}}(1) & \cdots & X_{N_{Ru}}(T) \end{pmatrix} \tag{33}$$

$X_i(t) = x_i^1(t) + x_i^2(t)\mathbf{i} + x_i^3(t)\mathbf{j} + x_i^4(t)\mathbf{k}, \ 1 \leq i \leq N_{Ru}, \ 1 \leq t \leq T$

The output matrix of the model in quaternion mode will be as follows,

$$Y = \begin{pmatrix} Y_1(1) & \cdots & Y_1(T) \\ \vdots & \ddots & \vdots \\ Y_{N_y}(1) & \cdots & Y_{N_y}(T) \end{pmatrix}, \tag{34}$$

$Y_i(t) = y_i^1(t) + y_i^2(t)\mathbf{i} + y_i^3(t)\mathbf{j} + y_i^4(t)\mathbf{k}, \ 1 \leq i \leq N_y, \ 1 \leq t \leq T$

Also, the matrix of input weights and the state network reservoir in the quaternion space are defined as follows:

$$W_{in} = \begin{pmatrix} W_{in,1}(1) & \cdots & W_{in,1}(N_U) \\ \vdots & \ddots & \vdots \\ W_{in,N_{Ru}}(1) & \cdots & W_{in,N_{Ru}}(N_U) \end{pmatrix},$$ (35)

$$W_{in,i}(n) = w_{in,i}^1(n) + w_{in,i}^2(n)\mathbf{i} + w_{in,i}^3(n)\mathbf{j} + w_{in,i}^4(n)\mathbf{k}, \ 1 \leq i \leq N_{Ru}, \ 1 \leq n \leq N_U$$

$$W = \begin{pmatrix} W_1(1) & \cdots & W_1(N_{Ru}) \\ \vdots & \ddots & \vdots \\ W_{N_{Ru}}(1) & \cdots & W_{N_{Ru}}(N_{Ru}) \end{pmatrix},$$ (36)

$$W_i(n) = w_i^1(n) + w_i^2(n)\mathbf{i} + w_i^3(n)\mathbf{j} + w_i^4(n)\mathbf{k}, \ 1 \leq i, n \leq N_{Ru}$$

In addition, the matrix of bilinear filter coefficients in the quaternion space will be as follows:

$$A = \begin{pmatrix} A_{1,1} & \cdots & A_{1,N_{Ru}+N_U} \\ \vdots & \ddots & \vdots \\ A_{N_y,1} & \cdots & A_{N_y,N_{Ru}+N_U} \end{pmatrix}$$ (37)

$$A_{i,j} = a_{i,j}^1 + a_{i,j}^2 \mathbf{i} + a_{i,j}^3 \mathbf{j} + a_{i,j}^4 \mathbf{k}, \ 1 \leq i \leq N_y, 1 \leq j \leq N_{Ru} + N_U$$

$$B^p = \begin{pmatrix} B_{1,1}^p & \cdots & B_{1,N_{Ru}+N_U}^p \\ \vdots & \ddots & \vdots \\ B_{(N-1),1}^p & \cdots & B_{(N-1),N_{Ru}+N_U}^p \end{pmatrix}$$

$$B_{i,j}^p = b_{i,j}^{p,1} + b_{i,j}^{p,2}\mathbf{i} + b_{i,j}^{p,3}\mathbf{j} + b_{i,j}^{p,4}\mathbf{k}, \ 1 \leq i \leq N - 1, \ 1 \leq j \leq N_{Ru} + N_U$$ (38)

$$B = [B^1, ..., B^{N_y}]$$

$$C = \begin{pmatrix} C_{1,1} & \cdots & C_{1,(N-1)} \\ \vdots & \ddots & \vdots \\ C_{N_y,1} & \cdots & C_{N_y,(N-1)} \end{pmatrix}$$ (39)

$$C_{i,j} = c_{i,j}^1 + c_{i,j}^2 \mathbf{i} + c_{i,j}^3 \mathbf{j} + c_{i,j}^4 \mathbf{k}, \ 1 \leq i \leq N_y, 1 \leq j \leq N - 1$$

Also, relation (8) of the echo state network will change in the form of the following relation in the quaternion state, so that the reservoir state matrix is obtained in this way,

$$X(t) = (1 - LR) * X(t-1) + LR * \tanh(W_{in} \otimes U(t) + W \otimes X(t-1))$$ (40)

In this regard, the tanh function is a split function and, as used in [14, 20, 21, 27, 50], will be applied to all dimensions of the quaternion number separately. In addition, the output of the Quaternion nonlinear echo state network after passing through the bilinear filter will be calculated using the following relation:

$$Y_p(t) = \sum_{i=1}^{N-1} C_{p,i} \otimes Y_p(t-i) + \sum_{i=1}^{N-1} \sum_{j=1}^{N_{Ru}+N_U} Y_p(t-i) \otimes B_{i,j}^p \otimes [X(t);U(t)]_j +$$
$$\sum_{i=1}^{N_{Ru}+N_U} A_{p,i} \otimes [X(t);U(t)]_i, \ 1 \leq p \leq N_y, \ N\text{:filter order} \tag{41}$$

The previous real multiplications are also replaced by the Hamilton multiplication in the following order:

$$\begin{aligned}
C_{p,i} \otimes Y_p(t-i) =\ & c_{p,i}^1 y_p^1(t-i) - c_{p,i}^2 y_p^2(t-i) - c_{p,i}^3 y_p^3(t-i) - c_{p,i}^4 y_p^4(t-i) \\
& + (c_{p,i}^1 y_p^2(t-i) + c_{p,i}^2 y_p^1(t-i) + c_{p,i}^3 y_p^4(t-i) - c_{p,i}^4 y_p^3(t-i))\mathbf{i} \\
& + (c_{p,i}^1 y_p^3(t-i) - c_{p,i}^2 y_p^4(t-i) + c_{p,i}^3 y_p^1(t-i) + c_{p,i}^4 y_p^2(t-i))\mathbf{j} \\
& + (c_{p,i}^1 y_p^4(t-i) + c_{p,i}^2 y_p^3(t-i) - c_{p,i}^3 y_p^2(t-i) + c_{p,i}^4 y_p^1(t-i))\mathbf{k}
\end{aligned} \tag{42}$$

$$\begin{aligned}
(Y_p(t-i) \otimes B_{i,j}^p) & \otimes [X(t);U(t)]_j = \\
& (y_p^1(t-i)b_{i,j}^{p,1}(t) - y_p^2(t-i)b_{i,j}^{p,2}(t) - y_p^3(t-i)b_{i,j}^{p,3}(t) - y_p^4(t-i)b_{i,j}^{p,4}(t) \\
& + (y_p^1(t-i)b_{i,j}^{p,2}(t) + y_p^2(t-i)b_{i,j}^{p,1}(t) + y_p^3(t-i)b_{i,j}^{p,4}(t) - y_p^4(t-i)b_{i,j}^{p,3}(t))\mathbf{i} \\
& + (y_p^1(t-i)b_{i,j}^{p,3}(t) - y_p^2(t-i)b_{i,j}^{p,4}(t) + y_p^3(t-i)b_{i,j}^{p,1}(t) + y_p^4(t-i)b_{i,j}^{p,2}(t))\mathbf{j} \\
& + (y_p^1(t-i)b_{i,j}^{p,4}(t) + y_p^2(t-i)b_{i,j}^{p,3}(t) - y_p^3(t-i)b_{i,j}^{p,2}(t) + y_p^4(t-i)b_{i,j}^{p,1}(t))\mathbf{k}) \\
& \otimes [X(t);U(t)]_j
\end{aligned} \tag{43}$$

$$\begin{aligned}
A_{p,i} \otimes X_i(t) =\ & a_{p,i}^1 x_i^1(t) - a_{p,i}^2 x_i^2(t) - a_{p,i}^3 x_i^3(t) - a_{p,i}^4 x_i^4(t) \\
& + (a_{p,i}^1 x_i^2(t) + a_{p,i}^2 x_i^1(t) + a_{p,i}^3 x_i^4(t) - a_{p,i}^4 x_i^3(t))\mathbf{i} \\
& + (a_{p,i}^1 x_i^3(t) - a_{p,i}^2 x_i^4(t) + a_{p,i}^3 x_i^1(t) + a_{p,i}^4 x_i^2(t))\mathbf{j} \\
& + (a_{p,i}^1 x_i^4(t) + a_{p,i}^2 x_i^3(t) - a_{p,i}^3 x_i^2(t) + a_{p,i}^4 x_i^1(t))\mathbf{k}
\end{aligned} \tag{44}$$

## 5.5. Quaternion Nonlinear Echo State Network by two-dimensional Principle Component Analysis

One of the limitations of the bilinear filter is the high dimensions of matrices A and B, which are completely dependent on the number of reservoir units. To reduce this complexity, as in the method proposed in [52,53], the dimensions of the input signal ($[X(t);U(t)]$) to the bilinear filter can be reduced by a method such as principal component analysis, which ultimately reduces the dimensions of matrices A and B. Principal component analysis is one of the methods that has been successful in processing information and reducing dimensions; But to apply this algorithm to quaternion matrices, they must be formed, which in turn eliminates the spatial correlation of adjacent data. This problem already existed when applying principal component analysis to image matrices. To solve this problem in the field of image, two-dimensional principal component analysis [60] has been proposed that does not require the process of converting an image or matrix to a vector. Since quaternion data are also three-dimensional matrices and require a similar image-to-vector conversion process to reduce the dimension by principal component analysis, two-dimensional principal component analysis can also be used to reduce the dimension in these data. The corresponding pseudocode is given in algorithm (1). In addition, in this algorithm, the implementation of quaternion nonlinear echo state network by two-dimensional principal component analysis method, the coefficients of which are obtained using genetic algorithm, is also shown. In this section, due to the stated limitations, the dimensions of the bilinear filter input matrix will be reduced by using the dimensional reduction method. In this case, the ESN will be as shown in Figure (10).

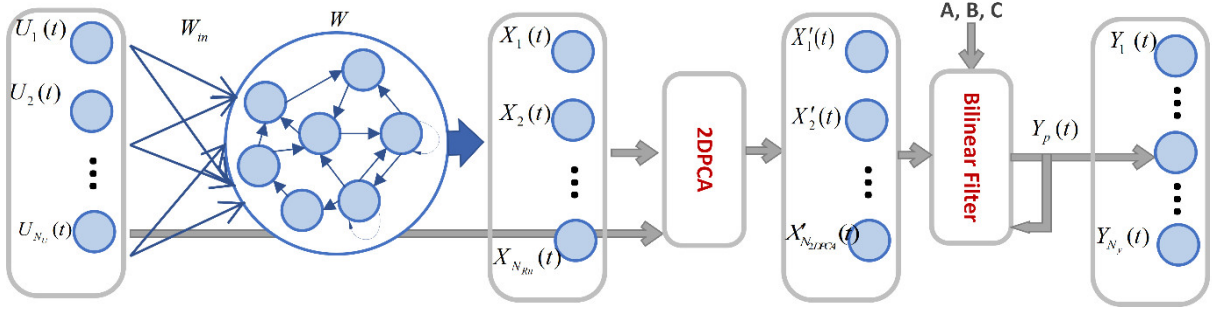

Figure 10. Quaternion nonlinear ESN with two-dimensianal principle component analysis

In addition, the previous relations from relation (32) to (36) will remain unchanged, except that relation (37), (38) and (40) will change as follows:

$$A = \begin{pmatrix} A_{1,1} & \cdots & A_{1,N_{2DPCA}} \\ \vdots & \ddots & \vdots \\ A_{N_y,1} & \cdots & A_{N_y,N_{2DPCA}} \end{pmatrix} \tag{45}$$

$$A_{i,j} = a_{i,j}^1 \mathbf{1} + a_{i,j}^2 \mathbf{i} + a_{i,j}^3 \mathbf{j} + a_{i,j}^4 \mathbf{k}, \ 1 \le i \le N_y, 1 \le j \le N_{2DPCA}$$

$$B^p = \begin{pmatrix} B_{1,1}^p & \cdots & B_{1,N_{2DPCA}}^p \\ \vdots & \ddots & \vdots \\ B_{(N-1),1}^p & \cdots & B_{(N-1),N_{2DPCA}}^p \end{pmatrix} \tag{46}$$

$$B_{i,j}^p = b_{i,j}^{p,1} \mathbf{1} + b_{i,j}^{p,2} \mathbf{i} + b_{i,j}^{p,3} \mathbf{j} + b_{i,j}^{p,4} \mathbf{k}, \ 1 \le i \le N-1, \ 1 \le j \le N_{2DPCA}$$

$$Y_p(t) = \sum_{i=1}^{N-1} C_{p,i} \otimes Y_p(t-i) + \sum_{i=1}^{N-1} \sum_{j=1}^{N_{2DPCA}} Y_p(t-i) \otimes B_{i,j}^p \otimes X_j'(t) + \sum_{i=1}^{N_{2DPCA}} A_{p,i} \otimes X_i'(t), \ 1 \le p \le N_y, \ N\text{:filter order} \tag{47}$$

Algorithm 1. Pseudo-code for speech emotion recognition using Quaternion nonlinear echo state network and two-dimensional principal component analysis with genetic algorithm

| **Algorithm** of RC Model with Bilinear Filter |
|---|
| 1: **Input:** *trainFeature, trainTarget, testFeature, testTarget* |
| 2: **Output:** *FinalPerformance* |
| 3: **for** $i = 1: foldNo$ **do** // for all folds |
| 4:     // all the executions have been repeated 10 times with 10 different RNG seeds |
| 5:     **for** $j = 1:10$ **do** |
| 6:         $rng(j-1)$ |
| 7:         $\theta_{Best} \leftarrow$ **GA-Optimization-method**(*trainFeature$_i$, trainTarget$_i$*); //Optimize the $\theta = [A, B, C, W_{in}, W]$ |
| 8: |
| 9:         $WAR_{i,j} \leftarrow$ **ObjectiveFunction**($\theta_{Best}$, *testFeature$_i$, testTarget$_i$*) |
| 10:     **end for** |
| 11: **end for** |
| 12: $FinalPerformance =$ **mean**($\{WAR_{i,j}\}$) |
| 13: **end function** |

**Algorithm** of Objective Function
1: **Input:** $A, B, C, W_{in}, W, U, Target$
2: **Output:** $fitness$
3: //Create an ESN
4: $X \leftarrow$ **ReservoirUnit**$(W_{in}, W, U)$ // compute reservoir unit output using $W_{in}, W$
5: $X' \leftarrow$ **2DPCA**$([X; U], N_{2DPCA})$ // reduce dimension by 2DPCA algorithm to $N_{2DPCA}$
6: //Compute ESN Output using Bilinear Filter
7: **for** $t = 1 : T$ **do** // for all samples belonging to X'
8:    **for** $p = 1 : N_y$ **do** // for all output dimension.
9: $$Y_p(t) = \sum_{i=1}^{N-1} C_{p,i} \otimes Y_p(t-i) + \sum_{i=1}^{N-1} \sum_{j=1}^{N_{2DPCA}} Y_p(t-i) \otimes B_{i,j}^p \otimes X'_j(t) + \sum_{i=1}^{N_{2DPCA}} A_{p,i} \otimes X'_i(t)$$
10:    **end for**
11: **end for**
12: **for** $utr_1 : utr_n$ **do** // for all utterances
13:    $P_{utr_i} =$ **mean**$(Y(k1:k2))$ // $k_1$:$k_2$ all samples belonging to $utr_i$
14:    $[E_1, \dots, E_{N_y}]=$**QMSE**$([D^1_{N_y \times 4}, \dots, D^{N_y}_{N_y \times 4}], P_{utr_i})$ // for all different class types $N_y$
15:    $Class_{utr_i} = \underset{1 \leq n \leq N_y}{\mathrm{argmax}}(e^{-E_n})$
16: **end for**
17: $fitness = \frac{\text{len}(Target==class)}{utrNo}$ // Target includes the output's label
18: **end function**

**Algorithm** of ReservoirUnit
1: **Input:** $W_{in}, W, U$
2: **Output:** $X$
3: **for** $t = 1 : T$ **do** // for all samples belonging to U
4:    $X_{1:N_{Ru}}(t) = (1 - LR) * X_{1:N_{Ru}}(t-1) + LR * \mathbf{tanh}(W_{in} \otimes U_{1:N_U}(t) + W \otimes X_{1:N_{Ru}}(t-1))$ // $N_{Ru}$: reservoir units No
5: **end for**
6: **end function**

**Algorithm** of 2dPCA
1: **Input:** $X, Dim$
2: **Output:** $X'$
3: $[m, n, Samples] =$ **size**$(X)$;
4: $XMean =$ **mean**$(X, 3)$; % Total mean of the training set
5: *% Computing covariance matrices*
6: $Gt =$ **zeros**$([n \ n])$;
7: **for** $i = 1: Samples$
8:   $Temp = X(:,:,i) - XMean$;
9:   $Gt = Gt + Temp' * Temp$;
10: **end for**
11: $Gt = Gt/Samples$;
12: *% Applying eigen-decomposition to Gt and returning transformation matrix*
13: $EigVect1 =$ **eig_decomp**$(Gt)$;
14: $EigVect = EigVect1(:,1:Dim)$;
15: *% Deriving reduced feature matrices*
16: **for** $i = 1: Samples$
17:   $X'(:,:,i) = X(:,:,i) * EigVect$;
18: **end for**
19: **end function**

In the proposed QNESN model, each quaternion property vector (both train and test data) will have dimensions of 554. Each dimension of a quaternion property is equivalent to one of the mid-term properties (mean, standard deviation, skewness, and kurtosis). These attributes apply to the extracted short-term features in the feature extraction section. In addition, in this algorithm, all parameters of the QNESN network, including the input weight matrix (relation 35), the reservoir weight (relation 36) and the matrix of linear filter coefficients (relations 45, 46 and 39) in the quaternion space will be defined as θ vector.

$$\theta = \left[ A, B, C, W_{in}, W \right] \tag{48}$$

This vector is optimized by the genetic algorithm. The initial population of the GA includes the random values in the range [-1 1]. This value is such that the echo state network retains its echo property, so the values given to the reservoir weight matrix (W) must be true in condition (14). In genetic algorithm, a QNESN network is designed with this population member to find the suitability of each member ($\theta_i$) of the generation (including ESN weights and bilinear filter coefficients) and its output is evaluated according to the training data compared to the desired output. For this evaluation, for each of the speech formats belonging to the training set, the average of all the frames related to that speech at the output of the network is obtained as follows, and then this average is scale to [0 1].

$$P_{utr_i} = \mathrm{mean}(Y(k_i : k_2)) \tag{49}$$

Where Y is the output of the QNESN network and k1 to k2 are the speech frames of $utr_i$, so the matrix $P_{utr_i}$ is a matrix with dimensions $N_y \times 4$ which is passed to the interval [0 1]. Then, using the QMSE algorithm presented in [16] the emotional class of that speech format, is obtained using the following relation,

$$[E_1, ..., E_{N_y}] = \mathrm{QMSE}\left( \left[ D^1_{N_y \times 4}, ..., D^{N_y}_{N_y \times 4} \right], P_{utr_i} \right) \tag{50}$$

$$Class_{utr_i} = \arg\max_{1 \leq n \leq N_y}(e^{-E_n}) \tag{51}$$

Where the set $[D^1_{N_y \times 4}, ..., D^{N_y}_{N_y \times 4}]$ contains labels for each output class defined as a quaternion matrix. For example, in the first-order quaternion matrix, only the first row has values of one, and the other objects are zero.

$$D^1_{N_y \times 4} = \begin{bmatrix} 1 & ... & 1 \\ \vdots & \ddots & \vdots \\ 0 & ... & 0 \end{bmatrix} \tag{52}$$

Then all the labels found by the network are compared with the desired output and the correct detection rate is measured. After completing the implementation of the genetic algorithm, and estimating the optimal values of weights and coefficients ($\theta_{best}$), using the best value obtained for these coefficients, a quaternion nonlinear echo state network is designed whose performance is tested by the data and pointed in the same way. The final performance will actually be the recognition rate of network on the experimental data set.

# 6. Experiments

All SER systems require an emotional database to evaluate their performance. Following the emotional databases properties used, will be described.

## 6.1. Emotional Datasets

Many emotional databases have been designed to test the performance of SER systems. In this study, three common databases including Berlin Database of Emotional Speech (EMODB) [61], Surrey Audio-Visual Expressed Emotion (SAVEE) [62], and the Interactive Emotional Dyadic Motion Capture (IEMOCAP) [63] were employed to evaluate the effectiveness of the proposed system, whose specifications are listed in Table 3.

Table 3. Specifications of emotional datasets

| Dataset | Language | Number of Speakers | Emotions | | | | | | | | Total |
|---|---|---|---|---|---|---|---|---|---|---|---|
| | | | Anger (A) | Disgust (D) | Fear (F) | Happiness (H) | Sad (SA) | Bore (B) | Surprise (SU) | Neutral (N) | |
| EMODB | German | 10 | 127 | 46 | 69 | 71 | 62 | 81 | - | 79 | 535 |
| SAVEE | English | 4 | 60 | 60 | 60 | 60 | 60 | - | 60 | 120 | 480 |
| IEMOCAP | English | 10 | 1103 | - | - | 1636 | 1084 | - | - | 1708 | 5531 |

## 6.2. Emotional Data Preparation

The results obtained using the new frame size (with 2W window length and K-frame shift window) and the application of mid-term properties are shown in Figures (11 and 12) for two different emotional databases. According to the presented results, the best value for window length was 40 frames and shift value was 10 frames for SAVEE database and window length was 40 frames and shift value was 30 frames for EMODB database. The feature vector extracted in this case has 2216 dimensions.

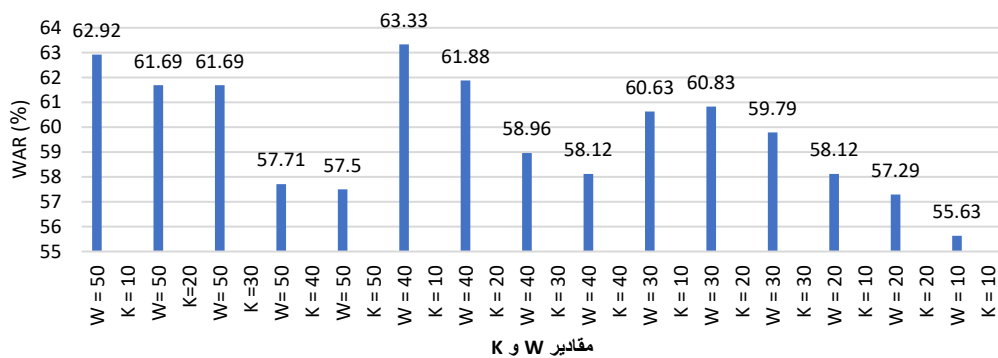

Figure 11. Recognition rate for windowing with different frames using echo state network (SAVEE)

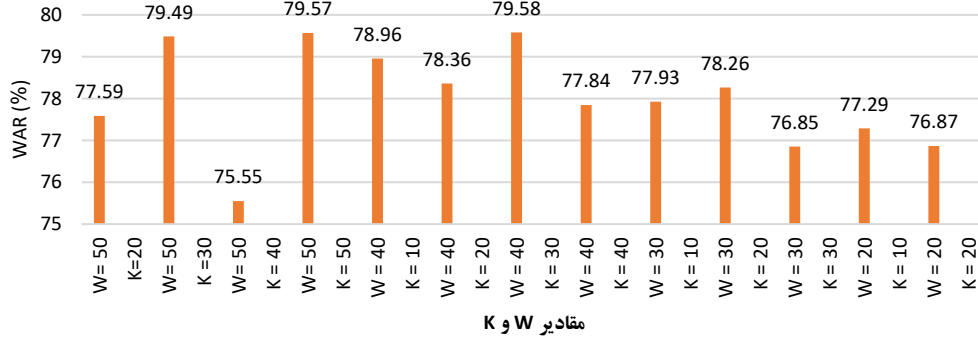

Figure 12. Recognition rate for windowing with different frames using echo state network (EMODB)

### 6.3. Network Hyper-parameter Regularization

With increasing depth and more complex structures of reservoir calculation models, optimization of the hyperparameters of these networks becomes more important. In this paper, various experiments have been performed to adjust the different values of C, SR, IS and LR hyperparameters that have a significant effect on the designed network performance. For convenience and speed of experiments, a simplified model of Figure (6) with a reservoir layer consisting of 100 neurons is provided.

In this experiment, different values of C are selected from the range $[10^{-6} ... 10^5]$ and the values related to each of the three parameters SR, IS and LR are selected from the range $[0… 1]$. All tests are performed on SAVEE database. According to the experiments performed, the best values obtained for these parameters are shown in Table 4.

Table 4. Optimal values of ESN hyperparameters

| $C = 100$ | $IS = 0.5$ | $SR = 0.1$ | $LR = 0.9$ |
|---|---|---|---|

### 6.4. QNESN Model Experiments

To evaluate and validate the performance of the proposed QNESN model, several experiments have been performed to detect speech emotion in the introduced emotion databases. In this study, the results of these experiments are presented in the leave one speaker out strategy (LOSO) through which many recent studies have evaluated their systems. According to this strategy, the emotional database is divided into two different parts, in which the training and experimental datasets are completely different, and the experimental database consists of only one speaker (i.e., a gender) who is not present in the training database. Therefore, network training and optimization is performed automatically on the sentences of the speakers of the training set and the recognition rate and network performance are tested on the speaker of the experimental set. After repeating the process for the total number of speakers, finally, the network performance for all speakers will be equal to the average recognition rate of different experiments.

In addition, this paper uses the weighted average recall (WAR) and non-weighted average recall (UAR) criteria to measure the recognition rate of experiments. The criterion for calling the weighted average or the micro average, as the recognition rate of the weighted average classes with the previous class probability, and the non-weighted average or the macro average calling, is the average recognition rate of all classes without weighting. The relationship between these two criteria is as follows,

$$Recall_i = \frac{TP_i}{TP_i + FN_i} \tag{53}$$

$$Weight_i = \frac{TP_i + FN_i}{N_e} \tag{54}$$

$$WAR = \sum_{i=1}^{M} Weight_i \times Recall_i \tag{55}$$

$$UAR = \frac{1}{M} \sum_{i=1}^{M} Recall_i \tag{56}$$

Where M is the number of emotions, $TP_i$ and $FN_i$, denoting true positives and false negatives for emotions i and $N_e$, respectively, represents the total number of samples of all emotions. To evaluate the performance of the proposed nonlinear model, the best value for the linear filter order (N) - which will have a significant effect on system performance - was first obtained during experiments on the SAVEE emotional database in the form of LOSO. In this experiment, a nonlinear echo state network (NESN) with 100 reservoir units and a bilinear filter with different levels has been used, the relationship of which has already been shown in (20). Bilinear filter coefficients (relations 23, 24 and 25) are obtained using genetic algorithm (the range of all coefficient values are considered in [-1,1]). The final results can be seen in Table (5). According to the obtained results, the 3-biline filter has a higher efficiency than other filters.

Table 5. The effect of the order value of a bilinear filter in the recognition rate of NESN network (SAVEE)

| N   | 2     | 3    | 4      | 5      |
|-----|-------|------|--------|--------|
| WAR | 46.5% | 54%  | 47.75% | 44.75% |

Experiments have also been performed to better and more accurately investigate the effect of the bilinear filter used in the output layer and the quaternion algebra in the ESN network. In this experiment, four different networks ESN, NESN, QESN and QNESN with filter order 3 and a number of different reservoir units are designed in which the values of weights and coefficients are optimized using genetic algorithm. In this optimization, which is done as LOSO, for each fold of the SAVEE dataset, a genetic algorithm is implemented based on the training data of that fold, and finally, using the optimal values of the obtained coefficients, the performance of the designed model is measured by those coefficients. The final efficiency will be equal to the average efficiency of different folds. Figure (13) shows the average and best value of the training cost functions of each generation of the genetic algorithm in successive iterations for one of the folds of the SAVEE database. In all experiments, each generation of the genetic algorithm has 500 members. Other characteristics of are shown in Table (6).

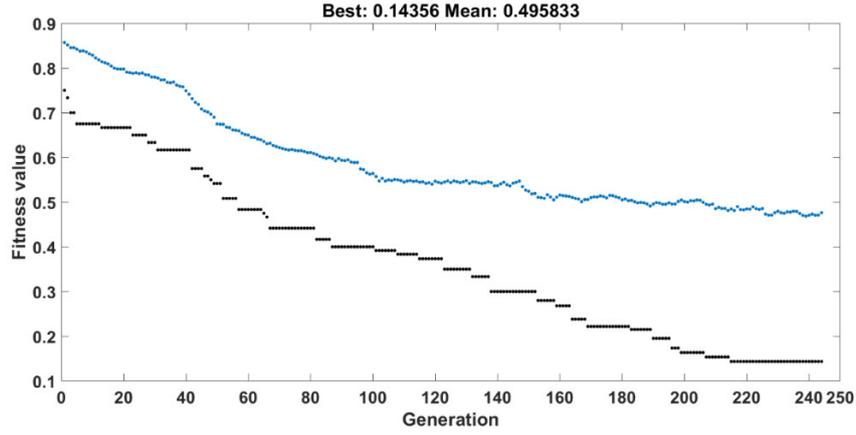

Figure 13. Mean and best value of training cost function in different iterations of QNESN parameter optimization process using genetic algorithm (SAVEE)

Table 6. The GA Specifications for optimizing ESN, NESN, QESN and QNESN network parameters

| Population | |
|---|---|
| Population type | Double vector |
| Population size | 500 |
| Creation Function | Uniform |
| Initial Population | [] |
| **Selection** | |
| Selection Function | Stochastic Uniform |
| **Reproduction** | |
| Elite Count | .05*Population size |
| Crossover Fraction | 0.8 |
| **Mutation** | |
| Mutation Function | Uniform |
| Rate | 0.01 |

| Crossover | |
|---|---|
| Crossover function | Scattered |
| **Migration** | |
| Direction | Forward |
| Fraction | 0.2 |
| Interval | 20 |
| **Constraint parameters** | |
| Nonlinear constrain algorithm | Augmented Lagrangian |
| Initial Penalty | 10 |
| Penalty Factor | 100 |
| **Stopping Criteria** | |
| Generations | 100*Number of Variables |
| Function Tolerance | 1e-6 |
| Constraint Tolerance | 1e-3 |

Table 7 shows the performance of five different networks for four different modes. In each part of the tests, the number of reservoir neurons, the size of the reservoir matrix, the size of the input signal at the output layer ($N_{Ru} + N_U$), the dimension reduction number, vector Θ and its size, recognition rates (UAR and WAR) and the average execution time for an audio file are shown. These experiments were performed in four different groups with different reservoirs. In all groups, the reservoir output will be given to the output layer after applying the reduction algorithm (PCA in real mode and 2dPCA in quaternion mode). To make the correct comparison in each group, the networks were tested under the same conditions. However, due to the fact that the network was a real or quaternion type, the dimensions of the input data will be different in the final layer. In addition, except for the first network, in which the weight of the output layer is calculated using the usual inverse matrix method, other networks use genetic algorithms to calculate weights and coefficients. Therefore, according to the design of each network, the vector θ and its size are different. The first network is the simple ESN in which the input weight matrix ($W_{in}$) according to relation (9) and the reservoir weight matrix (W) according to relation (10) are fixed and the output weight matrix ($W_{out}$) according to relation (11) is calculated using the inverse matrix method. In the second case, all weights of the input matrix, reservoir matrix, and output matrix of the ESN network are optimized using genetic algorithms. In this case the vector θ is obtained from the following relation,

$$\text{len}(\theta)_{ESN} = [(N_{Ru} \times N_U) + (N_{Ru} \times N_{Ru}) + (N_y \times N_{PCA})] \tag{57}$$

In the third network, which is NESN, only a bilinear filter is placed in the output layer, so not only the weights of the input matrix and reservoir matrix, but also the matrix of filter coefficients will be optimized using the genetic algorithm. Therefore, the vector length θ in this case will be longer than the normal ESN network and also due to the use of a bilinear filter in the output layer, it will have a higher execution time than the linear ESN network and its recognition rate is higher than the previous mode. The size of vector θ in this case is also obtained from the following relation:

$$\text{len}(\theta)_{NESN} = [(N_{Ru} \times N_U) + (N_{Ru} \times N_{Ru}) + (N_y \times N_{PCA}) + \\ (N_y \times (N-1) \times N_{PCA}) + (N_y \times (N-1))] \tag{58}$$

In the fourth case, the QESN quaternion network is tested. This network has a linear output layer, but all network weight matrices are defined as quaternions. Therefore, the length of θ will be longer than the previous three cases. The size of vector θ in this network is obtained from the following relation:

$$\text{len}(\theta)_{QESN} = [(N_{Ru} \times N_U \times 4) + (N_{Ru} \times N_{Ru} \times 4) + (N_y \times N_{2DPCA} \times 4)] \tag{59}$$

In the last case where the QNESN network is tested, a nonlinear ESN network with quaternion outputs is used, so the values of all matrix filter coefficients and network weights are optimized by the genetic algorithm, and the vector θ will be longer than in other cases.

$$\text{len}(\theta)_{QNESN} = [(N_{Ru} \times N_U \times 4) + (N_{Ru} \times N_{Ru} \times 4) + (N_y \times N_{2DPCA} \times 4) + \\ (N_y \times (N-1) \times N_{2DPCA} \times 4) + (N_y \times (N-1) \times 4)] \tag{60}$$

Table 7. Performance comparison (1) of ESN, NESN, QESN and QNESN networks with different configurations (SAVEE)

| $ESN - Type$ | Reservoir Unit No ($N_{Ru}$) | W matrix size | $N_{Ru} + N_U$ | $N_{2DPCA}$ | $\theta$ | $len(\theta)$ | WAR | UAR | Duration Time (s) |
|---|---|---|---|---|---|---|---|---|---|
| ESN | 25 | 25×25 | 25+2216+1=2242 | 200 | -- | -- | 57.29% | 53.10% | **0.3350** |
| ESN (GA) | 25 | 25×25 | 25+2216+1=2242 | 200 | $[W_{in}, W, W_{out}]$ | 57450 | 58.00% | 54.76% | 0.3229 |
| NESN | 25 | 25×25 | 25+2216+1=2242 | 200 | $[A, B, C, W_{in}, W]$ | 60264 | 59.03% | 55.70% | 0.3203 |
| QESN | 25 | 25×25 | 25+554+1=580 | 200 | $[W_{in}, W, W_{out}]$ | 63600 | 61.80% | 60.18% | 0.2923 |
| QNESN | 25 | 25×25 | 25+554+1=580 | 200 | $[A, B, C, W_{in}, W]$ | **74856** | **62.22%** | **60.60%** | 0.2661 |
| ESN | 50 | 50×50 | 50+2216+1=2266 | 400 | -- | -- | 61.46% | 58.10% | 0.3195 |
| ESN (GA) | 50 | 50×50 | 50+2216+1=2266 | 400 | $[W_{in}, W, W_{out}]$ | 116150 | 62.10% | 59.71% | 0.3239 |
| NESN | 50 | 50×50 | 50+2216+1=2266 | 400 | $[A, B, C, W_{in}, W]$ | 121764 | 62.46% | 60.05% | 0.3254 |
| QESN | 50 | 50×50 | 50+554+1=605 | 400 | $[W_{in}, W, W_{out}]$ | 132200 | 64.95% | 63.05% | **0.5241** |
| QNESN | 50 | 50×50 | 50+554+1=605 | 400 | $[A, B, C, W_{in}, W]$ | **154656** | **65.49%** | **63.60%** | 0.4832 |
| ESN | 100 | 100×100 | 100+2216+1=2317 | 400 | -- | -- | 61.46% | 58.10% | 0.3366 |
| ESN (GA) | 100 | 100×100 | 100+2216+1=2317 | 400 | $[W_{in}, W, W_{out}]$ | 234500 | 62.27% | 59.88% | 0.3390 |
| NESN | 100 | 100×100 | 100+2216+1=2317 | 400 | $[A, B, C, W_{in}, W]$ | 240114 | 62.86% | 60.43% | 0.3583 |
| QESN | 100 | 100×100 | 100+554+1=655 | 400 | $[W_{in}, W, W_{out}]$ | 273200 | 65.56% | 63.65% | **1.0175** |
| QNESN | 100 | 100×100 | 100+554+1=655 | 400 | $[A, B, C, W_{in}, W]$ | **295656** | **66.48%** | **64.57%** | 1.0003 |
| ESN | 200 | 200×200 | 200+2216+1=2417 | 500 | -- | -- | 61.46% | 57.98% | 0.3973 |
| ESN (GA) | 200 | 200×200 | 200+2216+1=2417 | 500 | $[W_{in}, W, W_{out}]$ | 486900 | 62.21% | 59.69% | 0.3950 |
| NESN | 200 | 200×200 | 200+2216+1=2417 | 500 | $[A, B, C, W_{in}, W]$ | 493914 | 63.31% | 60.73% | 0.4110 |
| QESN | 200 | 200×200 | 200+554+1=755 | 500 | $[W_{in}, W, W_{out}]$ | 618000 | 65.52% | 63.64% | **2.2917** |
| QNESN | 200 | 200×200 | 200+554+1=755 | 500 | $[A, B, C, W_{in}, W]$ | **646056** | **66.67%** | **65.77%** | 2.2383 |

According to the descriptions, due to time-consuming Hamiltonian multiplication operation, in each group, the highest execution time is usually related to quaternion networks and the lowest execution time is related to the simple ESN network. In each group, due to the properties of the bilinear filter, the use of nonlinear network has produced better results compared to the linear network, and this filter has done the classification of features at the output more accurately. The use of quaternion numbers has also produced more accurate results than a real-valued network. In this table, in each group, the maximum values related to the length θ, the maximum execution time, and the maximum values of UAR and WAR are specified. Thus, the highest amount of WAR obtained using QESN and QNESN networks with 100 and 200 reservoirs is 65.56% and 66.67%, respectively, and the highest recognition rates are obtained for ESN and NESN networks with 100 and 200 reservoirs, which are 62.27% and 63.31%, respectively. According to the results, QNESN network is more efficient than other networks, due to the nonlinear structure used in the output, the use of Hamiltonian multiplication to discover relationships between different data and optimization of network coefficients using GA.

In a more detailed comparison, shown in Table 8, the real and quaternion networks with equivalent reservoir sizes are in the same group. In the second category, the real networks ESN and NESN with 200-reservoir with their equivalent quaternion networks (QESN and QNESN) with 50-quaternion reservoir (equivalent to 200 real reservoirs), in terms of memory consumption (θ vector length) and recognition rate (WAR) have been compared. In this group, QNESN and QESN quaternion networks with less memory consumption produce better results compared to real networks. In the third group, with 400 real reservoir units and 100 quaternion reservoir units, QESN and QNESN quaternion networks have been able to produce better results compared to real networks with less memory consumption (θ length). In addition, in the last group (800 real units and 200 quaternion units), quaternion networks have been able to have this advantage over real networks in both higher accuracy and lower memory consumption. In fact, in the comparison made in this table, the efficiency of using quaternion algebra and bilinear filters in increasing the accuracy and precision of the SER system with less memory consumption compared to its real equivalent networks can be seen. Of course, this increase in accuracy has been further improved by increasing the size of the reservoir compared to real networks, and at lower dimensions (group 1 in Table 8), real networks have performed better than quaternion networks (see Figures 14 and 15 for further comparison). In addition, in the last row of the table, the effect of using dimensional reduction using the two-dimensional principal component analysis algorithm is clearly visible. It can be concluded that the use of 2DPCA reduction algorithm has a positive effect on the overall results of the SER system.

Table 9 shows the amount of memory consumed in the process of optimizing the parameters of ESN, NESN, QESN and NQESN networks. This table shows the length of each chromosome (according to the size of the θ vector) as well as the amount of memory consumed by the genetic algorithm per population of 500 (one generation). In addition, the specifications of the system on which the tests were performed are also visible. The relationship between the amount of memory consumed during optimization and the final accuracy of the generated SER system is well illustrated in this table. As can be seen, quaternion networks with less memory consumption during the optimization process have ultimately produced higher accuracy (see Figure 16 for further comparison).

To further evaluate the performance of the proposed algorithm, all previous experiments on the EMODB database have been performed (Table 10). Also, the best results obtained for the EMODB, SAVEE and IEMOCAP databases are shown in comparison with other work done so far and in the same conditions in Tables 11 to 13. According to the results presented in these two tables, the proposed architecture in this section has a higher efficiency compared to many recent works done on SAVEE, EMODB and IEMOCAP databases.

Table 8. Comparison of the performance (2) of ESN, NESN, QESN and QNESN networks with different configurations (SAVEE)

| $ESN-Type$ | Reservoir Unit No ($N_{Ru}$) | W matrix size | $N_{Ru} + N_U$ | $N_{2DPCA}$ | $\theta$ | $len(\theta)$ | WAR | UAR | Duration Time (s) |
|---|---|---|---|---|---|---|---|---|---|
| ESN | 100 | 100×100 | 100+2216+1=2317 | 400 | -- | -- | 61.46% | 58.10% | 0.3366 |
| ESN (GA) | 100 | 100×100 | 100+2216+1=2317 | 400 | $[W_{in}, W, W_{out}]$ | 234500 | 62.27% | 59.88% | 0.3390 |
| NESN | 100 | 100×100 | 100+2216+1=2317 | 400 | $[A, B, C, W_{in}, W]$ | **240114** | **62.86%** | 60.43% | **0.3583** |
| QESN | 25 | 25×25 | 25+554+1=580 | 200 | $[W_{in}, W, W_{out}]$ | 63600 | 61.80% | 60.18% | 0.2923 |
| QNESN | 25 | 25×25 | 25+554+1=580 | 200 | $[A, B, C, W_{in}, W]$ | 74856 | 62.22% | **60.60%** | 0.2661 |
| ESN | 200 | 200×200 | 200+2216+1=2417 | 500 | -- | -- | 61.46% | 57.98% | 0.3973 |
| ESN (GA) | 200 | 200×200 | 200+2216+1=2417 | 500 | $[W_{in}, W, W_{out}]$ | 486900 | 62.21% | 59.69% | 0.3950 |
| NESN | 200 | 200×200 | 200+2216+1=2417 | 500 | $[A, B, C, W_{in}, W]$ | 493914 | 63.31% | 60.73% | 0.4110 |
| QESN | 50 | 50×50 | 50+554+1=605 | 400 | $[W_{in}, W, W_{out}]$ | 132200 | 64.95% | 63.05% | **0.5241** |
| QNESN | 50 | 50×50 | 50+554+1=605 | 400 | $[A, B, C, W_{in}, W]$ | 154656 | **65.49%** | **63.60%** | 0.4832 |
| ESN | 400 | 400×400 | 400+2216+1=2617 | 500 | -- | -- | 61.46% | 57.98% | 0.4845 |
| ESN (GA) | 400 | 400×400 | 400+2216+1=2617 | 500 | $[W_{in}, W, W_{out}]$ | 1210300 | 62.67% | 60.12% | 0.4946 |
| NESN | 400 | 400×400 | 400+2216+1=2617 | 500 | $[A, B, C, W_{in}, W]$ | 1217314 | 63.89% | 61.25% | 0.4901 |
| QESN | 100 | 100×100 | 100+554+1=655 | 400 | $[W_{in}, W, W_{out}]$ | 273200 | 65.56% | 63.65% | **1.0175** |
| QNESN | 100 | 100×100 | 100+554+1=655 | 400 | $[A, B, C, W_{in}, W]$ | 295656 | **66.48%** | **64.57%** | 1.0003 |
| ESN | 800 | 800×800 | 800+2216+1=3017 | 600 | -- | -- | 62.08% | 58.57% | 0.7514 |
| ESN (GA) | 800 | 800×800 | 800+2216+1=3017 | 600 | $[W_{in}, W, W_{out}]$ | 3057800 | 63.12% | 60.54% | 0.7400 |
| NESN | 800 | 800×800 | 800+2216+1=3017 | 600 | $[A, B, C, W_{in}, W]$ | 3066214 | 64.55% | 61.90% | 0.6857 |
| QESN | 200 | 200×200 | 200+554+1=755 | 500 | $[W_{in}, W, W_{out}]$ | 618000 | 65.52% | 63.64% | **2.2917** |
| QNESN | 200 | 200×200 | 200+554+1=755 | 500 | $[A, B, C, W_{in}, W]$ | 646056 | **66.67%** | **65.77%** | 2.2383 |
| QNESN | 200 | 200×200 | 200+554+1=755 | -- | $[A, B, C, W_{in}, W]$ | 667476 | 65.16% | 64.28% | 1.0124 |

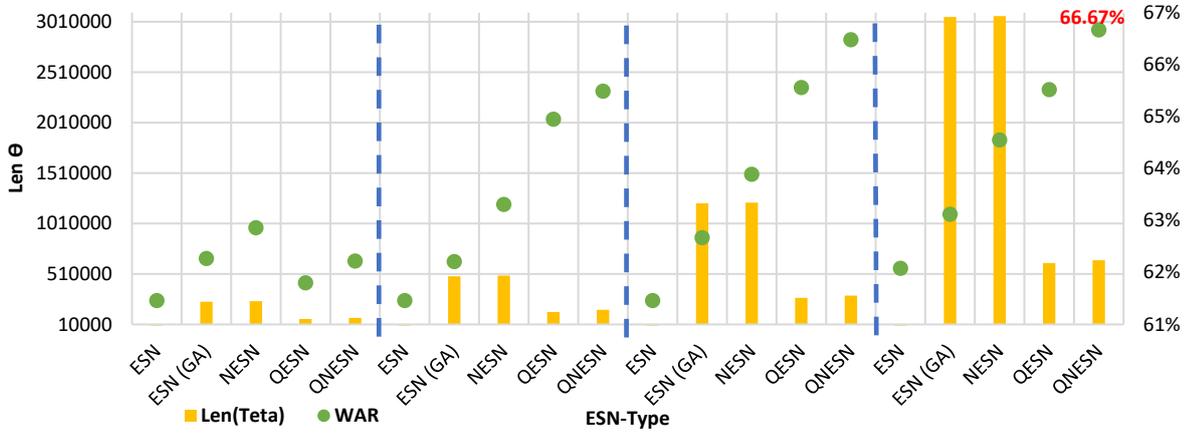

Figure 14. Comparison of different ESN types from Table (8) in terms of performance (WAR) and θ length (SAVEE)

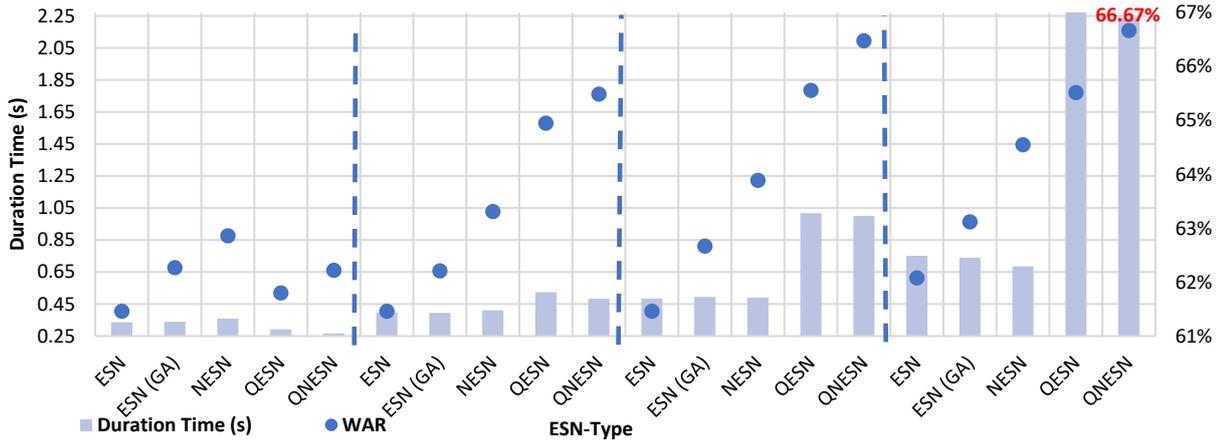

Figure 15. Comparison of different ESN types from Table (8) in terms of performance (WAR) and duration time (SAVEE)

Table 9. Comparison of memory consumption of genetic algorithm in ESN, NESN, QESN and QNESN network optimization process with different configurations (SAVEE)

| $ESN-Type$ | Reservoir Unit No ($N_{Ru}$) | $len(\theta)$ (double vector) | Chromosome Length | GA Memory Consumption (One Generation) | CPU Specification | WAR |
|---|---|---|---|---|---|---|
| ESN | 100 | -- | -- | -- | Core™ i3-4160-3.60GHz 8.00 GB | 61.46% |
| ESN (GA) | 100 | 234500 | 1.87 MB | 938 MB | Xeon(R) E7-2699-2.30GHz 10.00 GB | 62.27% |
| **NESN** | 100 | **240114** | **1.92 MB** | **960.45 MB** | Xeon(R) E7-2699-2.30GHz 10.00 GB | **62.86%** |
| QESN | 25 | 63600 | 0.5088 MB | 254.4 MB | Xeon(R) E7-2699-2.30GHz 10.00 GB | 61.80% |
| QNESN | 25 | 74856 | 0.5988 MB | 299.42 MB | Xeon(R) E7-2699-2.30GHz 10.00 GB | 62.22% |
| ESN | 200 | -- | -- | -- | Core™ i3-4160-3.60GHz 8.00 GB | 61.46% |
| ESN (GA) | 200 | 486900 | 3.89 MB | 1.945 GB | Core™ i3-4160-3.60GHz 8.00 GB | 62.21% |
| **NESN** | 200 | **493914** | **3.95 MB** | **1.976 GB** | Core™ i3-4160-3.60GHz 8.00 GB | 63.31% |
| QESN | 50 | 132200 | 1.05 MB | 528.8 MB | Core™ i7-M640-2.80GHz 8.00 GB | 64.95% |
| QNESN | 50 | 154656 | 1.23 MB | 618.62 MB | Core™ i7-M640-2.80GHz 8.00 GB | **65.49%** |
| ESN | 400 | -- | -- | -- | Core™ i3-4160-3.60GHz 8.00 GB | 61.46% |
| ESN (GA) | 400 | 1210300 | 9.68 MB | 4.84 GB | Core™ i7-3720QM-2.60GHz 16.00 GB | 62.67% |
| **NESN** | 400 | **1217314** | **9.73 MB** | **4.86 GB** | Core™ i7-3720QM-2.60GHz 16.00 GB | 63.89% |
| QESN | 100 | 273200 | 2.18 MB | 1.09 GB | Core™ i3-4160-3.60GHz 8.00 GB | 65.56% |
| QNESN | 100 | 295656 | 2.36 MB | 1.18 GB | Core™ i3-4160-3.60GHz 8.00 GB | **66.48%** |
| ESN | 800 | -- | -- | -- | Core™ i3-4160-3.60GHz 8.00 GB | 62.08% |
| ESN (GA) | 800 | 3057800 | 24.46 MB | 12.23 GB | Core™ i5-6400-2.70GHz 32.00 GB | 63.12% |
| **NESN** | 800 | **3066214** | **24.52 MB** | **12.26 GB** | Core™ i5-6400-2.70GHz 32.00 GB | 64.55% |
| QESN | 200 | 618000 | 4.94 MB | 2.47 GB | Core™ i3-4160-3.60GHz 8.00 GB | 65.52% |
| QNESN | 200 | 646056 | 5.16 MB | 2.58 GB | Core™ i3-4160-3.60GHz 8.00 GB | **66.67%** |

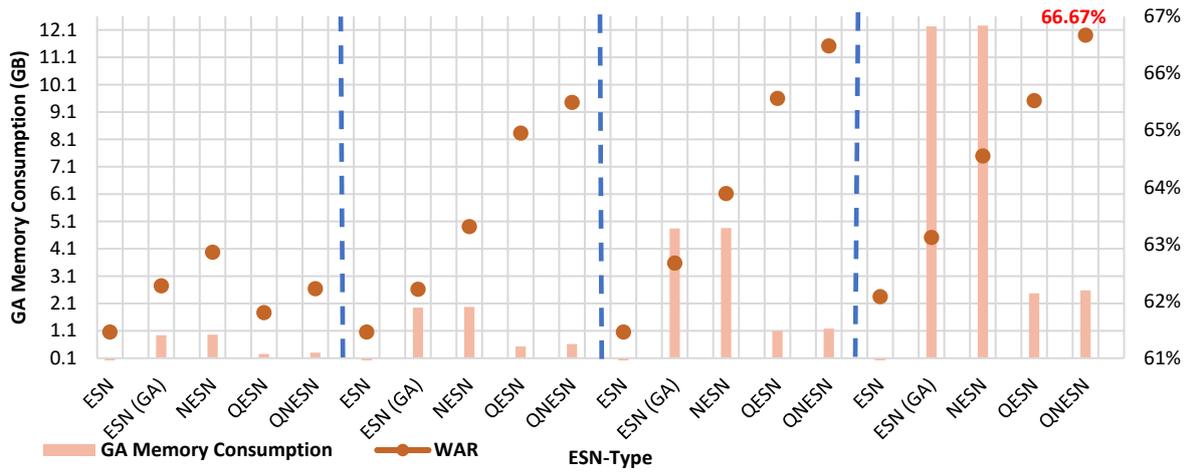

Figure 16. Comparison of different ESN networks from Table (9) in terms of WAR and memory consumption of genetic algorithm (SAVEE)

Table 10. Comparison of the performance of ESN, NESN, QESN and QNESN networks with different configurations (EMODB)

| $ESN-Type$ | Reservoir Unit No ($N_{Ru}$) | W matrix size | $N_{Ru} + N_U$ | $N_{2DPCA}$ | $\theta$ | $len(\theta)$ | WAR | UAR | Duration Time (s) |
|---|---|---|---|---|---|---|---|---|---|
| ESN | 100 | 100×100 | 100+2216+1=2317 | 400 | -- | -- | 78.45% | 78.19% | 0.42 |
| ESN (GA) | 100 | 100×100 | 100+2216+1=2317 | 400 | $[W_{in}, W, W_{out}]$ | 234500 | 79.97% | 79.26% | 0.4162 |
| NESN | 100 | 100×100 | 100+2216+1=2317 | 400 | $[A, B, C, W_{in}, W]$ | **240114** | **80.69%** | **79.85%** | 0.4256 |
| QESN | 25 | 25×25 | 25+554+1=580 | 200 | $[W_{in}, W, W_{out}]$ | 63600 | 80.27% | 78.79% | 0.6387 |
| QNESN | 25 | 25×25 | 25+554+1=580 | 200 | $[A, B, C, W_{in}, W]$ | 74856 | 80.52% | 79.21% | **0.6786** |
| ESN | 200 | 200×200 | 200+2216+1=2417 | 500 | -- | -- | 77.86% | 77.80% | 0.4895 |
| ESN (GA) | 200 | 200×200 | 200+2216+1=2417 | 500 | $[W_{in}, W, W_{out}]$ | 486900 | 79.57% | 78.61% | 0.4931 |
| NESN | 200 | 200×200 | 200+2216+1=2417 | 500 | $[A, B, C, W_{in}, W]$ | **493914** | 80.61% | 79.71% | 0.4888 |
| QESN | 50 | 50×50 | 50+554+1=605 | 400 | $[W_{in}, W, W_{out}]$ | 132200 | 82.93% | 81.35% | 1.1551 |
| QNESN | 50 | 50×50 | 50+554+1=605 | 400 | $[A, B, C, W_{in}, W]$ | 154656 | **83.48%** | **81.89%** | 1.2408 |
| ESN | 400 | 400×400 | 400+2216+1=2617 | 500 | -- | -- | 77.86% | 77.86% | 0.6513 |
| ESN (GA) | 400 | 400×400 | 400+2216+1=2617 | 500 | $[W_{in}, W, W_{out}]$ | 1210300 | 80.00% | 79.07% | 0.6404 |
| NESN | 400 | 400×400 | 400+2216+1=2617 | 500 | $[A, B, C, W_{in}, W]$ | **1217314** | 81.13% | 80.29% | 0.5864 |
| QESN | 100 | 100×100 | 100+554+1=655 | 400 | $[W_{in}, W, W_{out}]$ | 273200 | 83.53% | 81.96% | 1.9484 |
| QNESN | 100 | 100×100 | 100+554+1=655 | 400 | $[A, B, C, W_{in}, W]$ | 295656 | **84.45%** | **82.88%** | **2.49** |
| ESN | 800 | 800×800 | 800+2216+1=3017 | 600 | -- | -- | 79.42% | 78.99% | 0.802 |
| ESN (GA) | 800 | 800×800 | 800+2216+1=3017 | 600 | $[W_{in}, W, W_{out}]$ | 3057800 | 80.76% | 80.32% | 0.8186 |
| NESN | 800 | 800×800 | 800+2216+1=3017 | 600 | $[A, B, C, W_{in}, W]$ | **3066214** | 82.67% | 82.22% | 0.8585 |
| QESN | 200 | 200×200 | 200+554+1=755 | 500 | $[W_{in}, W, W_{out}]$ | 618000 | 83.83% | 83.37% | 4.8026 |
| QNESN | 200 | 200×200 | 200+554+1=755 | 500 | $[A, B, C, W_{in}, W]$ | 646056 | **85.39%** | **84.92%** | 5.7045 |

Table 11. EMODB recognition rates (%) compared to the QNESN model

| Ref | WAR | UAR | Ref | WAR | UAR | Ref | WAR | UAR |
|---|---|---|---|---|---|---|---|---|
| Haider 2020, [45] | N/A | 76.90 | Rintala 2020, [76] | 78.50 | N/A | Tzinis 2018 [93] | 82.40 | N/A |
| Zhao 2020, [38] | N/A | 79.70 | Wang 2020, [41] | 79.20 | N/A | Kalinli 2016 [95] | 82.70 | N/A |
| Chen 2018, [64] | N/A | 82.82 | Hassan 2012, [77] | 79.50 | N/A | Hou 2020 [48] | 82.80 | 83.30 |
| Sidorov 2016, [65] | 72.00 | N/A | Daneshfar 2020, [78] | 79.94 | 76.81 | Daneshfar 2020, [79] | 82.82 | N/A |
| Aghajani 2020, [66] | 72.10 | N/A | Zao 2014, [80] | 80.10 | N/A | Zhang 2021 [47] | 83.70 | N/A |
| Li 2021, [67] | 72.19 | N/A | Bhargava 2013, [81] | 80.60 | N/A | Yi 2019 [39] | 83.74 | N/A |
| Yüncü 2014, [68] | 72.30 | N/A | Badshah 2019, [82] | 80.79 | N/A | Deb 2018 [73] | 83.80 | N/A |
| Khan 2017, [32] | 72.34 | N/A | Zhang 2013, [83] | 80.85 | N/A | Yi 2020 [97] | 84.49 | 83.31 |
| Sinith 2015, [69] | 73.75 | N/A | Wu 2011, [84] | 81.30 | N/A | Tawari 2010 [98] | 84.50 | N/A |
| Deb 2017, [71] | 73.90 | N/A | Sun 2015 [30] | 81.50 | N/A | Özseven 2018 [99] | 84.50 | N/A |
| Deb 2016, [70] | 74.40 | N/A | Sun 2015 [85] | 81.74 | N/A | Özseven 2019 [100] | 84.62 | N/A |
| Tao 2016, [74] | 74.46 | N/A | Xu 2015 [86] | 81.80 | N/A | Deb 2017, [72] | 85.10 | N/A |
| Kadiri 2015 [31] | 75.22 | N/A | Vieira 2020, [49] | 81.80 | N/A | Meng 2019 [101] | 85.32 | N/A |
| Shirani 2016, [75] | 76.12 | N/A | Man-Wai 2016 [87] | 81.86 | N/A | **Proposed** | **85.39** | **84.92** |
| Bashirpour 2016, [34] | 76.60 | N/A | Stuhlsatz 2011 [88] | 81.90 | N/A | Sajjad 2020 [127] | 85.57 | N/A |
| Sugan 2020, [42] | 77.08 | N/A | Wen 2017 [36] | 82.32 | N/A | Chen 2020, [44] | 85.61 | N/A |
| Luengo 2010, [35] | 78.30 | N/A | Lotfidereshgi 2017 [89] | 82.35 | N/A | Singh 2020 [46] | 86.36 | N/A |
| | | | Sun 2017 [29] | 82.40 | N/A | Er 2020 [126] | 90.21 | N/A |

Table 12. SAVEE recognition rates (%) compared to the QNESN model

| Ref | WAR | UAR |
|---|---|---|
| Haider 2020 [45] | N/A | 42.4 |
| Papakostas 2017 [102] | 44.00 | N/A |
| Liu 2018 [103] | 44.18 | N/A |
| Noroozi and Marjanovic 2017 [104] | 45.51 | N/A |
| Vásquez-Correa 2016 [106] | 47.30 | N/A |
| Sun 2015 [30] | 50.00 | N/A |
| Sun 2017 [29] | 51.46 | 49.33 |
| Wen 2017 [36] | 53.60 | N/A |
| Tzinis 2018 [93] | 54.00 | 53.80 |
| Sugan 2020, [42] | 55.83 | N/A |
| Noroozi and Sapiński 2017 [105] | 56.07 | N/A |
| Sinith 2015 [69] | 57.50 | N/A |
| Sun 2015 [85] | 58.76 | N/A |
| Daneshfar 2020, [79] | 59.38 | 55.00 |
| Zhang 2021 [47] | 60.16 | N/A |
| Daneshfar 2020, [78] | 60.79 | N/A |
| Nguyen 2020 [107] | 62.00 | N/A |
| Jiang 2019 [108] | 62.49 | 59.40 |
| Wang 2020 [41] | 66.20 | 81.8 |
| **Proposed** | **66.67** | **65.77** |
| Farooq 2020 [109] | 66.90 | N/A |

Table 13. IEMOCAP recognition rates (%) compared to the QNESN model

| Ref | WAR | UAR | Ref | WAR | UAR | Ref | WAR | UAR |
|---|---|---|---|---|---|---|---|---|
| Latif 2019 [110] | N/A | 60.23 | Hou 2020 [48] | 62.80 | 63.80 | Zhao 2020 [38] | 65.20 | 68.00 |
| Zong 2018 [111] | N/A | 64.80 | Tzinis 2018 [93] | 63.00 | 65.20 | Daneshfar 2020, [78] | 65.71 | 65.73 |
| Latif 2020 [40] | N/A | 66.70 | Li 2015 [116] | 63.20 | N/A | **Proposed** | **66.32** | **63.11** |
| Kwon 2020 [112] | N/A | 73.01 | Vieira 2020, [49] | 63.2 | N/A | Deb 2018, [73] | 66.80 | N/A |
| Ghosh 2016 [113] | 52.82 | N/A | Jiang 2019 [108] | 64.00 | N/A | Yi 2019 [39] | 66.80 | 62.83 |
| Xie 2019 [114] | 54.00 | N/A | Tzinis 2017 [94] | 64.16 | N/A | Yi 2020 [97] | 66.92 | 64.51 |
| Li 2020 [43] | 58.62 | 59.91 | Deb 2017, [71] | 64.20 | N/A | Liu 2018 [103] | 67.10 | 66.20 |
| Zhao 2018 [118] | 59.70 | 60.10 | Chen 2018 [64] | 64.20 | N/A | Yeh 2020 | 69.00 | 70.10 |
| Huang 2018 [119] | 60.40 | N/A | Han 2018 [121] | 64.20 | 65.70 | Sun 2020 [92] | 71.50 | N/A |
| Li 2021 [67] | 60.83 | N/A | Issa 2020 [122] | 64.30 | N/A | Li 2018 [115] | 71.75 | N/A |
| Xia 2015 [15] | 60.90 | 62.40 | Etienne 2018 [123] | 64.50 | 61.70 | Fan 2020 | 73.02 | 65.86 |
| Zhao 2019 [117] | 61.90 | N/A | Fayek 2017 [124] | 64.80 | 60.90 | Daneshfar 2020, [79] | 74.80 | N/A |
| Mao 2019 [120] | 62.28 | 58.02 | Shirani 2016 [75] | 65.20 | N/A | | | |

# 7. Analysis of the QNESN model
## 7.1. Comparison of the Proposed Method with Recent Researches

Tables 14 to 16 show the confusion matrices for the best results obtained so far on the EMODB, SAVEE, and IEMOCAP emotion databases. According to the EMODB database confusion matrices, angry emotion with 91.11% accuracy has a higher recognition rate than other emotions. In addition, sad emotion had the highest similarity to bore emotion (compared to other emotions) due to similar vocal characteristics and glottal waveform (90.95% to 6.79%). Also, angry emotion has the highest similarity to happy emotion among other emotions (91.11% to 4.15%). This similarity is also due to the similarity of the prosodic features (both happy and angry emotions have more energy than other emotions) and the glottal waveform for these two emotions. In addition, in the case of the SAVEE and IEMOCAP databases confusion matrices, the most similar emotion to the happy emotion is angry (60.36% to 10.66%) and (61.29% to 15.74%).

Compared to other proposed architectures, the method [41] using features based on wavelet packet analysis, could not distinguish seven different emotions from each other in the EMODB database (low recognition rate of fear emotion, 55.07% and neutral 56.96%, compared to the recognition rate of 79.09% and 90.01% in the QNESN method). This also true for the SAVEE and IEMOCAP emotional databases. In addition, in the method presented by [44], due to the use of membership algorithm to select appropriate features, the recognition rate of some emotions such as neutral (96.20% vs. 90.01%) and fear (84% vs. 0.09) 79%) is higher than the QNESN method and the recognition rate of some emotions such as disgust (63.20% versus 87.01%), bore (80.6% versus 81.24%) and anger (74% versus 11 / 91%) is less than QNESN method. These results indicate that the effect of glottal wave signal, linear perceptual coefficient and Gabor filter-based features on emotions such as anger, bore and disgust are more than the emotion of neutral and fear. Also, the results of the proposed QNESN network in recognizing the emotions of the SAVEE emotion database and also in recognizing the emotions of the IEMOCAP emotion database have been remarkable compared to many of the results obtained so far and in the same experimental conditions. However, in the architecture presented in this paper, due to the use of Hamiltonian multiplication, the response time of the algorithm is longer than the similar real network. The maximum response time obtained for the EMODB database according to Table (10) in this algorithm is 5.745 seconds, which is smaller than the execution time reported by [125] on the same database (8.23 seconds). The maximum execution time on this emotional base in the architecture presented using real networks is even less than this value (0.802 seconds). The main reason for the high speed of the proposed emotion recognition system compared to other architectures is the use of fast classifier based on echo state network.

Table 14. EMODB confusion matrix using QNESN

| | | **Predicted** | | | | | | |
|---|---|---|---|---|---|---|---|---|
| | | F | D | H | B | N | SA | A |
| **Real** | Fear | 79.09 | 0.75 | 3.11 | 0 | 9.01 | 3.01 | 5.02 |
| | Disgust | 5.11 | 87.01 | 0.22 | 0 | 0.17 | 2.28 | 5.21 |
| | Happiness | 5.04 | 9.4 | 75.01 | 0 | 0.56 | 0 | 10 |
| | Bore | 0 | 3.88 | 0 | 81.24 | 5.01 | 9.87 | 0.01 |
| | Neutral | 1.05 | 3.44 | 0.25 | 4.89 | 90.01 | 0.11 | 0.24 |
| | Sad | 0 | 0 | 0.25 | 6.79 | 2 | 90.95 | 0 |
| | Anger | 0.78 | 1.1 | 4.15 | 1.2 | 1.55 | 0.21 | 91.11 |

Table 15. SAVEE confusion matrix using QNESN

|  | | Predicted | | | | | |
|---|---|---|---|---|---|---|---|
| | | A | D | F | H | N | SA | SU |
| Real | A | 60.36 | 7.24 | 4.02 | 10.66 | 5.65 | 3.14 | 8.9 |
| | D | 7.85 | 62.94 | 5.9 | 7 | 8.41 | 4.7 | 3.21 |
| | F | 8.89 | 4.88 | 62.66 | 9.21 | 7.87 | 5.27 | 1.2 |
| | H | 6.14 | 2.1 | 3.07 | 78.63 | 3.74 | 1 | 5.22 |
| | N | 4.25 | 2.29 | 8.77 | 7.98 | 61.51 | 10.24 | 4.95 |
| | SA | 5.02 | 2.2 | 7 | 8.94 | 7.11 | 69.72 | 0 |
| | SU | 8.4 | 5 | 6.9 | 5.12 | 9.96 | 0 | 64.58 |

In following, table (17) shows the recognition rate of the QNESN network for each of the ten folds of the EMODB database. According to the obtained values, the recognition rate in these experiments has increased well for the emotions expressed by each speaker and with each gender. In fact, the presented architecture has well recognized both the emotions expressed by the female speaker (86.7%) and the emotions expressed by the male speaker (84.08%). In these experiments, the use of SFCC features is very helpful in recognizing the same emotions expressed by different speakers and categorizing them as the same emotion, and finally increasing the average and final recognition rate by increasing the recognition rate in each fold.

Table 16. IEMOCAP confusion matrix using QNESN

| | Angry | Happiness | Neutral | Sad |
|---|---|---|---|---|
| Anger | 61.29 | 15.74 | 11.98 | 11 |
| Happiness | 13.55 | 64.56 | 11.46 | 10.44 |
| Neutral | 7.95 | 12.26 | 65.02 | 14.76 |
| Sad | 10.8 | 19 | 8.62 | 61.58 |

Table 17. Recogntion rate (%) on EMODB for different speakers and genders using QNESN model

| fold number (gender in the test set) | WAR |
|---|---|
| 1 (m) | 82.99 |
| 2 (f) | 86.88 |
| 3 (f) | 89.95 |
| 4 (m) | 80.14 |
| 5 (m) | 80.84 |
| 6 (m) | 90.79 |
| 7 (f) | 87.69 |
| 8 (f) | 80.17 |
| 9 (m) | 85.66 |
| 10 (f) | 88.83 |
| **Mean (m)** | **84.08** |
| **Mean (f)** | **86.70** |

Figure 17 compares the method presented in this paper in terms of the execution time (a) and the audio file response time (for one second of audio file) (b) with recent references that have reported these parameters. Meanwhile, in the model presented by [97], the execution time of the proposed learning algorithm for the generative adversarial network is 0.07 seconds in each iteration using the graphical processing unit (GPU). For the general implementation of this algorithm, 800 different iteration are required. Finally, the total execution time required for this algorithm is 56 seconds. In the model provided by Air [126], the best result provided using the DenseNet201 network requires 364 seconds of training time on the EMODB dataset using the GPU. In the architecture presented by [127], the total execution time of the proposed algorithm on the EMODB dataset is 5396 seconds using GPU. In a study done by [128] using the GPU, the total execution time of the learning algorithm was 1850 seconds and the response time to one second of the audio file (emotion recognition only) in the experimental data set was 0.4 seconds. In the model presented by [46], the best classifier execution time on the processor with specifications (Corei5-8GB RAM) is reported to be 2299 seconds. In the architecture presented by [44], the response time for the best results obtained using the multiple random forest classifier was 0.07 seconds (unfortunately CPU specifications were not reported). Ali Bakhshi [129] performed all his tests on a processor with specifications (Intel Corei7-4GB RAM) and reported the total execution time on 170 speech utterances from the EMODB database, 5 minutes (the average length of each speech utterance was 2.5 seconds). Finally, the response time for 1 second of the audio file was approximately 0.71 seconds. In the research presented by [112], the response time to 1 second of the speech utterance of the EMODB dataset was reported to be 0.4 seconds using a processor with specifications (Corei7). In the research presented by [125], the execution time of the proposed algorithm on the EMODB data set using a processor (Core E5-48GB RAM) is reported to be 8.23 seconds. Among these, the shortest execution time among all recent models has been obtained by [97] and also the shortest response time to 1 second of the input file, among the recent models presented by [44]. Execution time using the pQPSO dimension reduction algorithm and the GMM statistical classifier (previously published by the authors [79]) for each fold of the EMODB database, on a processor with Intel Core i3–8 GB RAM specifications, was 0.75 hours, which is finally, it lasted 7.5 hours on 10 different folds (Figure 17, part A). The use of GMM statistical classifier with high components (128 components) and the use of the objective function, based on the development dataset recognition rate using GMM, have been the most important reasons for the high execution time of this algorithm. Despite the high execution time, this algorithm has a short response time (Figure 17, part b). This time was 0.6 seconds for 1 second of each speech utterance. The execution time was 7.64 hours using the pQPSO dimension reduction algorithm and the GEBF statistical classifier (previously published by the same authors [78]) on the same processor (Figure 17, Part A). The use of GMM statistical classifier with high components (1024 components) has been the most important reason for the high execution time of this algorithm. Despite the high execution time, this algorithm has a better response time (Figure 17, part b). This time was 1.1 seconds for each second of each speech utterance. However, the time required for the QNESN classifier has reached 5 hours on the entire dataset and on the same processor due to parameter optimization using GA and Hamiltonian multiplication, and the response time is 1.52 seconds. The reason for increasing the response time of QNESN compared to the methods [78] and [79] is the use of Hamiltonian multiplication in MATLAB environment, which is time consuming. Compared to other references, since the implementation and execution environment of the programs are not the same and also experiments have been performed with different processors and in many references, GPUs have been used, different execution and response times have been obtained.

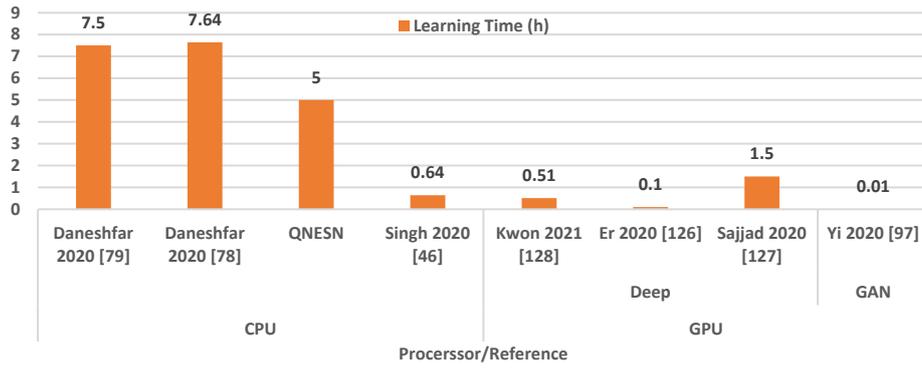

(a)

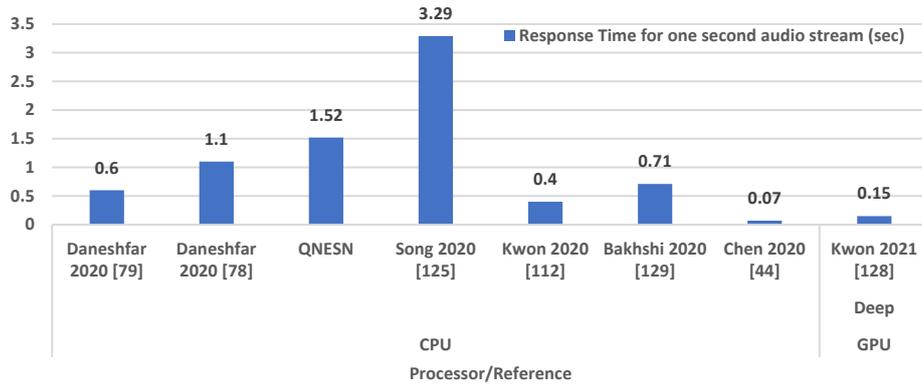

(b)

Figure 17. Comparison of (a) the execution time and (b) the response time of the propoed method with other references (EMODB)

## 7.2. Statistical Analysis of Recognition Rates

Figure 18 compares the method presented in this paper with other methods presented by the authors of this paper ([78-79]). As it turns out, the highest mean value among the three different emotional databases is related to EMODB and the lowest value is related to SAVEE database. Due to the small number of samples of the training set compared to the samples of the experimental set in the SAVEE dataset (compared to the EMODB and IEMOCAP datasets), a significant decrease in the recognition rate is evident in this database.

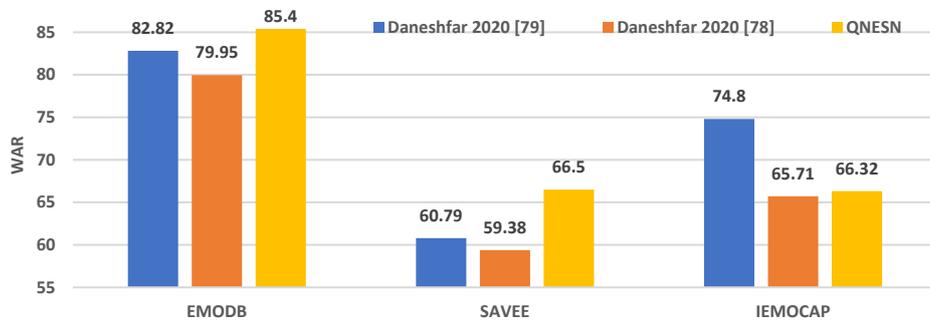

Figure 18. Comparison of recognition rate (WAR) of the proposed method using measure of central tendency in comparison with other methods proposed by the authors

These results indicate that the mid-term features used in the QNESN method perform better on artificial speech formats (EMODB and SAVEE datasets). However, these features on IEMOCAP speech formats, which include both artificial conversations and improvised scenarios, have not yielded the desired results. In addition, the short-term features used in [79] are suitable for longer speech utterances (the length of the speech utterances for the part of IEMOCAP used in these experiments is $4.73 \pm 0.18$ seconds), while the mid-term features used in the QNESN method are suitable for utterances with shorter average length (EMODB 2 to 3 seconds and SAVEE $3.85 \pm 0.33$ seconds). Therefore, it can be concluded that the mid-term features, and the QNESN method will produce higher recognition rates in datasets with shorter lengths and fewer samples, but detect and differentiate more emotions (7 different emotions). However, short-term features and [79] in the dataset with longer length and more samples but for recognizing fewer emotions (4 emotions) will be better.

### 7.3. Statistical Analysis of Different Emotion Recognition Rates

Figure (19) shows the recognition rates of different emotions for the proposed method and for different emotion sets. According to section a, in the EMODB, angry, sad, and neutral emotions are identified at a higher rate than other emotions due to the large number of samples. This is also evident in section c for the neutral emotion in IEMOCAP. In addition, sad and angry emotions with fewer samples in

IEMOCAP had lower recognition rates. According to section (d), the emotions of bore, happiness, sadness and fear with mid-term features and QNESN method had higher recognition rates and neutral, angry, disgust and surprised emotions with short-term features and using method [79] had better results. Emotions with instantaneous changes (such as angry and surprised emotions) appear to have higher recognition rates using short-term features and the GMM statistical classifier. Also, the emotions of bore, happiness, sadness and fear are better modeled as time series and mid-term features and its dynamics are recognized by QNESN classifier with higher recognition rate. However, the method presented in this paper does not recognize the emotion of disgust better than other emotions. Because disgust speech has fewer vocal features than other emotions and therefore has few examples in the emotional data collections, then it is more complex than other emotions and requires better classifiers and the use of more relevant features to recognition.

In general, it seems that the emotional features proposed in this paper have improved the average rate of recognition of different emotions and have well differentiated different emotions from each other. In other words, the proposed method, on average, has been able to distinguish different emotions (in the case of the same speakers and different speakers) from each other, and also to consider well the same emotions expressed by the same speakers and different speakers.

### 8. Discussion

In this paper, a new method for designing SER systems based on the QNESN network as a classifier is proposed, which makes the final architecture more accurate than many recently published SER systems. This study uses the features of GBFB, SFCC, PLPC and glottal waveform of speech input signal, which has been effective in reducing redundancy and differentiation between different emotions. In this research, an efficient set of long-term statistics (mean, standard deviation, skewness and kurtosis), are considered as different dimensions of quaternion numbers, each of which evaluates the speech signal from a specific perspective. The nonlinear classifier presented in this paper is a dynamic scalable model that has a high memory capacity for various features. According to experiments, features that pass through a nonlinear output produce better results than features that pass through a linear output structure. In fact, by using the nonlinear filter in the output layer, ESN can classify the input data with less error. This progress has been achieved without increasing the complexity of the filter coefficient training process using the 2dPCA technique. Such nonlinear structures are better suited to time series problems as well as large nonlinear problems such as emotion recognition than simple structures such as ESN and ELM networks. In addition, the results show that the proposed model has a strong dynamic behavior and therefore will be useful for modeling high-dimensional data at different times and in different hierarchies. Another reason for the high recognition rate of this model for emotional databases is the ability to detect external relationships between the multidimensional features of a sequence as well as their internal latent structural dependencies using Hamiltonian multiplication and lower parameters. The use of quaternion algebra causes the high-dimensional feature to be encapsulated and the size of the reservoir matrix to be one-fourth (relative to real numbers). Because in the case of real numbers, if $N_{Ru}$ is the number of reservoir units, the size of the reservoir matrix will be $N_{Ru} \times N_{Ru}$, but using quaternion algebra this size will be reduced to $\frac{N_{Ru}}{4} \times \frac{N_{Ru}}{4} \times 4$. Although quaternion algebra and quaternion multiplication are powerful tools in reducing the amount of memory consumed by the state network, the execution time and optimization of the network coefficients due to the Hamiltonian multiplication structure will be longer than the real state and this is the biggest limitation of using these numbers. Therefore, the coefficient optimization phase can be performed on parallel computers or GPUs to increase the speed of the QNESN coefficient optimization process. Although the proposed architecture has a higher execution time compared to ESN and ELM in real data mode, it ultimately runs much faster than other deeper models, and will run less than 60 minutes for samples with less than 1000 dimensions, which is a small amount of time. In the future, the performance of this structure can be investigated in the multi-corpus state and also for other emotional databases. In addition, the results reported in this study are based on emotional databases in noise-free environments, while real databases are instantaneous and noisy, in which the proposed frameworks need to be further developed, which will be another research path for the future work of the authors.

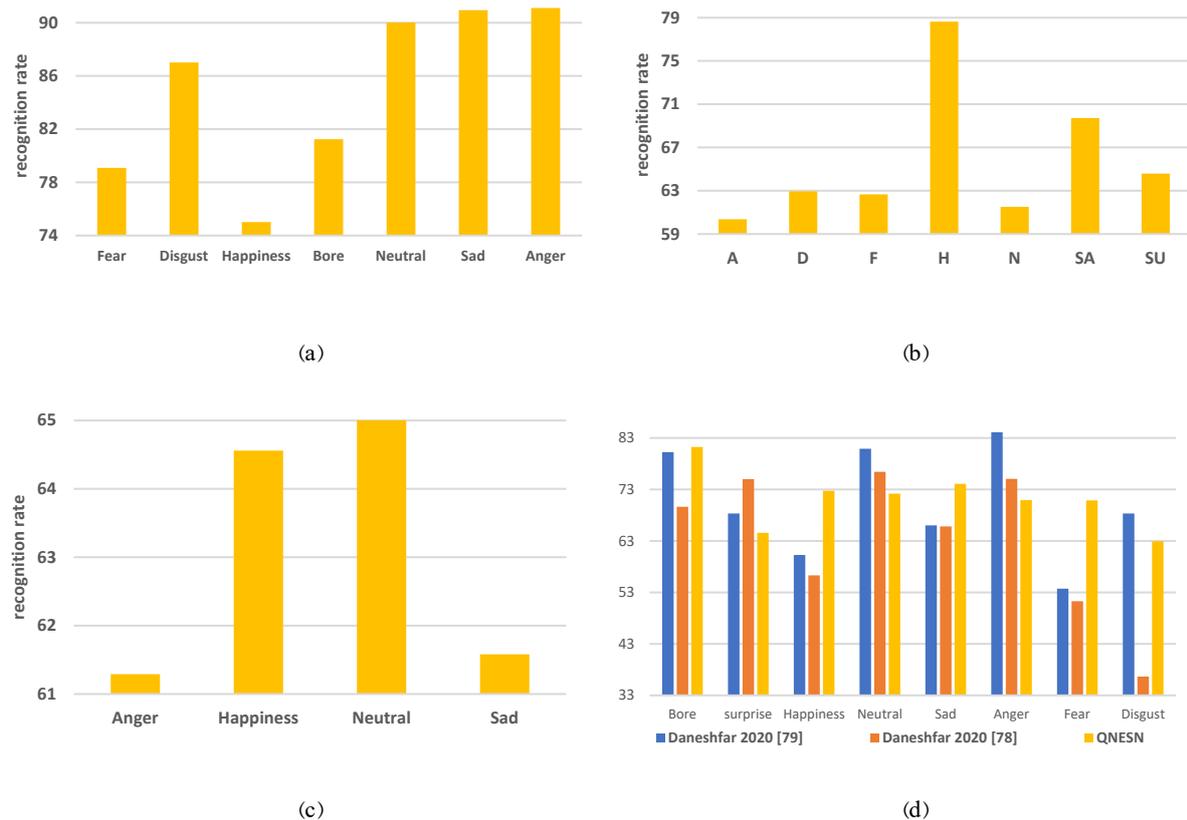

Figure 19. Comparison of recogniton rate (WAR) of different emotions a) EMODB b) SAVEE c) IEMOCAP d) All data

## 9. Conclusion

In this paper, a nonlinear model of a quaternion echo state network is proposed to speech emotion recognition. The use of a multidimensional bilinear filter in the output layer of this network has enabled the system to have a good ability to transfer nonlinear features from the reservoir layer to the output and to classify the input data with less error. Also, with the help of this filter, the proposed model has made good use of the higher order statistics of reservoir signals. In this model, the use of the 2dPCA technique before the bilinear filter has caused that only a part of the main components of the reservoir state has been used by this filter, and this has prevented it from increasing its complexity. In addition, in the proposed model, quaternion algebra is used to reduce the amount of memory consumed by the reservoir. These numbers provide a more compact representation of multidimensional speech features, resulting in a higher capacity to use high-dimensional data. Also, by using Hamilton multiplications, the external relations between the features and their internal latent structural dependencies are extracted with less parameters. The proposed architecture for recognizing speech emotion has been evaluated on three different emotional databases with different features. According to the results presented in different stages, the proposed nonlinear quaternion structure has better results than the simple structure of the echo state network. The results of this study not only indicate an increase in QNESN network performance, but also describe and introduce it as a new signal processing tool.

## References

[1] Lukoševičius, M. and Jaeger, H., 2009. Reservoir computing approaches to recurrent neural network training. *Computer Science Review*, *3*(3), pp.127-149.

[2] Soh, H. and Demiris, Y., 2015. Learning assistance by demonstration: Smart mobility with shared control and paired haptic controllers. *Journal of Human-Robot Interaction*, *4*(3), pp.76-100.

[3] Xu, M. and Han, M., 2016. Adaptive elastic echo state network for multivariate time series prediction. *IEEE transactions on cybernetics*, *46*(10), pp.2173-2183.

[4] Ma, Q., Shen, L. and Cottrell, G.W., 2020. DeePr-ESN: A deep projection-encoding echo-state network. *Information Sciences*, *511*, pp.152-171.

[5] Shen, L., Chen, J., Zeng, Z., Yang, J. and Jin, J., 2018. A novel echo state network for multivariate and nonlinear time series prediction. *Applied Soft Computing*, *62*, pp.524-535.

[6] Massar, M. and Massar, S., 2013. Mean-field theory of echo state networks. *Physical Review E*, *87*(4), p.042809.

[7] Jaeger, H., 2002. Short term memory in echo state networks. *GMD Report*, *1*, p.52.

[8] Wainrib, G. and Galtier, M.N., 2016. A local Echo State Property through the largest Lyapunov exponent. *Neural Networks*, *76*, pp.39-45.

[9] Yildiz, I.B., Jaeger, H. and Kiebel, S.J., 2012. Re-visiting the echo state property. *Neural networks*, *35*, pp.1-9.

[10] Lukoševičius, M., Jaeger, H. and Schrauwen, B., 2012. Reservoir computing trends. *KI-Künstliche Intelligenz*, *26*(4), pp.365-371

[11] Jaeger, H., 2001. The "echo state" approach to analysing and training recurrent neural networks-with an erratum note. *Bonn, Germany: German National Research Center for Information Technology GMD Technical Report*, *148*(34), p.13.

[12] Parcollet T, Zhang Y, Morchid M, Trabelsi C, Linarès G, de Mori R, Bengio Y (2018) Quaternion convolutional neural networks for end-to-end automatic speech recognition. In: Interspeech 2018, 19th Annual conference of the international speech communication association, Hyderabad, India, 2–6 September 2018, pp 22–26. https://doi.org/10.21437/Interspeech.2018-1898

[13] Parcollet, T., Morchid, M. and Linares, G., 2017, December. Deep quaternion neural networks for spoken language understanding. In *2017 IEEE Automatic Speech Recognition and Understanding Workshop (ASRU)* (pp. 504-511). IEEE.

[14] Parcollet, T., Morchid, M., Linarès, G. and De Mori, R., 2019, May. Bidirectional quaternion long short-term memory recurrent neural networks for speech recognition. In *ICASSP 2019-2019 IEEE International Conference on Acoustics, Speech and Signal Processing (ICASSP)* (pp. 8519-8523). IEEE.

[15] Xia, R. and Liu, Y., 2015. A multi-task learning framework for emotion recognition using 2D continuous space. *IEEE Transactions on affective computing*, *8*(1), pp.3-14.

[16] Arena, P., Fortuna, L., Re, R. and Xibilia, M.G., 1993, May. On the capability of neural networks with complex neurons in complex valued functions approximation. In *1993 IEEE International Symposium on Circuits and Systems* (pp. 2168-2171). IEEE.

[17] Schädler, M.R. and Kollmeier, B., 2015. Separable spectro-temporal Gabor filter bank features: Reducing the complexity of robust features for automatic speech recognition. *The Journal of the Acoustical Society of America*, *137*(4), pp.2047-2059.

[18] Aneeja, G., and B. Yegnanarayana. "Single frequency filtering approach for discriminating speech and nonspeech." *IEEE/ACM Transactions on Audio, Speech, and Language Processing* 23, no. 4 (2015): 705-717

[19] Parcollet, Titouan, Mohamed Morchid, and Georges Linarès. "Quaternion convolutional neural networks for heterogeneous image processing." In *ICASSP 2019-2019 IEEE International Conference on Acoustics, Speech and Signal Processing (ICASSP)*, pp. 8514-8518. IEEE, 2019

[20] Qiu, X., Parcollet, T., Ravanelli, M., Lane, N. and Morchid, M., 2020. Quaternion Neural Networks for Multi-channel Distant Speech Recognition. *arXiv preprint arXiv:2005.08566*.

[21] Gaudet, C.J. and Maida, A.S., 2018, July. Deep quaternion networks. In *2018 International Joint Conference on Neural Networks (IJCNN)* (pp. 1-8). IEEE.

[22] Takahashi K, Isaka A, Fudaba T, Hashimoto M (2017) Remarks on quaternion neural network-based controller trained by feedback error learning. In: 2017 IEEE/SICE International symposium on system integration (SII), pp 875–880

[23] Ogawa T (2016) Neural network inversion for multilayer quaternion neural networks. Comput Technol Appl 7:73–82

[24] Bayro-Corrochano E, Lechuga-Gutiérrez L, Garza-Burgos M (2018) Geometric techniques for robotics and hmi: Interpolation and haptics in conformal geometric algebra and control using quaternion spike neural networks. Robot Auton Syst 104:72–84

[25] De Leo S, Rotelli P (1997) Local hypercomplex analyticity. arXiv preprint arXiv:9703002 [funct-an]

[26] Nitta T (1995) A quaternary version of the back-propagation algorithm. In: IEEE International conference on neural networks, 1995. Proceedings, vol 5. IEEE, pp 2753–2756

[27] Zhu, X., Xu, Y., Xu, H. and Chen, C., 2018. Quaternion convolutional neural networks. In *Proceedings of the European Conference on Computer Vision (ECCV)* (pp. 631-647).

[28] Wu, J., Xu, L., Wu, F., Kong, Y., Senhadji, L. and Shu, H., 2020. Deep octonion networks. *Neurocomputing*

[29] Sun, Y. and Wen, G., 2017. Ensemble softmax regression model for speech emotion recognition. *Multimedia Tools and Applications*, *76*(6), pp.8305-8328.

[30] Sun, Y. and Wen, G., 2015. Emotion recognition using semi-supervised feature selection with speaker normalization. *International Journal of Speech Technology*, *18*(3), pp.317-331.

[31] Kadiri, S.R., Gangamohan, P., Gangashetty, S.V. and Yegnanarayana, B., 2015. Analysis of excitation source features of speech for emotion recognition. In *Sixteenth Annual Conference of the International Speech Communication Association*.

[32] Khan, A. and Roy, U.K., 2017, March. Emotion recognition using prosodie and spectral features of speech and Naïve Bayes Classifier. In *2017 international conference on wireless communications, signal processing and networking (WiSPNET)* (pp. 1017-1021). IEEE.

[33] Deb, S. and Dandapat, S., (a) 2017, March. Exploration of phase information for speech emotion classification. In *2017 Twenty-third National Conference on Communications (NCC)* (pp. 1-5). IEEE.

[34] Bashirpour, M. and Geravanchizadeh, M., 2016. Speech emotion recognition based on power normalized cepstral coefficients in noisy conditions. *Iranian Journal of Electrical and Electronic Engineering*, *12*(3), pp.197-205.

[35] Luengo, I., Navas, E. and Hernáez, I., 2010. Feature analysis and evaluation for automatic emotion identification in speech. *IEEE Transactions on Multimedia*, *12*(6), pp.490-501.

[36] Wen, G., Li, H., Huang, J., Li, D. and Xun, E., 2017. Random deep belief networks for recognizing emotions from speech signals. *Computational intelligence and neuroscience*, *2017*.

[37] Vasquez-Correa, J.C., Arias-Vergara, T., Orozco-Arroyave, J.R., Vargas-Bonilla, J.F. and Noeth, E., 2016, October. Wavelet-based time-frequency representations for automatic recognition of emotions from speech. In *Speech Communication; 12. ITG Symposium* (pp. 1-5). VDE.

[38] Zhao, H., Xiao, Y. and Zhang, Z., 2020. Robust Semisupervised Generative Adversarial Networks for Speech Emotion Recognition via Distribution Smoothness. *IEEE Access*, *8*, pp.106889-106900.

[39] Yi, L. and Mak, M.W., 2019, November. Adversarial data augmentation network for speech emotion recognition. In *2019 Asia-Pacific Signal and Information Processing Association Annual Summit and Conference (APSIPA ASC)* (pp. 529-534). IEEE.

[40] Latif, S., Rana, R., Khalifa, S., Jurdak, R., Epps, J. and Schuller, B.W., 2020. Multi-task semi-supervised adversarial autoencoding for speech emotion recognition. *IEEE Transactions on Affective Computing*.

[41] Wang, K., Su, G., Liu, L. and Wang, S., 2020. Wavelet packet analysis for speaker-independent emotion recognition. *Neurocomputing*.

[42] Sugan, N., Srinivas, N.S.S., Kumar, L.S., Nath, M.K. and Kanhe, A., 2020. Speech emotion recognition using cepstral features extracted with novel triangular filter banks based on bark and ERB frequency scales. *Digital Signal Processing*, p.102763.

[43] Li, H., Tu, M., Huang, J., Narayanan, S. and Georgiou, P., 2020, May. Speaker-Invariant Affective Representation Learning via Adversarial Training. In *ICASSP 2020-2020 IEEE International Conference on Acoustics, Speech and Signal Processing (ICASSP)* (pp. 7144-7148). IEEE.

[44] Chen, L., Su, W., Feng, Y., Wu, M., She, J. and Hirota, K., 2020. Two-layer fuzzy multiple random forest for speech emotion recognition in human-robot interaction. *Information Sciences*, *509*, pp.150-163.

[45] Haider, F., Pollak, S., Albert, P. and Luz, S., 2020. Emotion recognition in low-resource settings: An evaluation of automatic feature selection methods. *Computer Speech & Language*, p.101119

[46] Singh, R., Puri, H., Aggarwal, N. and Gupta, V., 2020. An Efficient Language-Independent Acoustic Emotion Classification System. *Arabian Journal for Science and Engineering*, *45*(4), pp.3111-3121.

[47] Zhang, Z., 2021. Speech feature selection and emotion recognition based on weighted binary cuckoo search. *Alexandria Engineering Journal*, *60*(1), pp.1499-1507.

[48] Hou, M., Li, J. and Lu, G., 2020. A supervised non-negative matrix factorization model for speech emotion recognition. *Speech Communication*, *124*, pp.13-20.

[49] Vieira, V., Coelho, R. and de Assis, F.M., 2020. Hilbert–Huang–Hurst-based non-linear acoustic feature vector for emotion classification with stochastic models and learning systems. *IET Signal Processing*, *14*(8), pp.522-532.

[50] Shang, F. and Hirose, A., 2013. Quaternion neural-network-based PolSAR land classification in Poincare-sphere-parameter space. *IEEE Transactions on Geoscience and Remote Sensing*, *52*(9), pp.5693-5703

[51] Gallicchio, C. and Micheli, A., 2017. Echo state property of deep reservoir computing networks. *Cognitive Computation*, *9*(3), pp.337-350.

[52] Boccato, L., Lopes, A., Attux, R. and Von Zuben, F.J., 2011, July. An echo state network architecture based on Volterra filtering and PCA with application to the channel equalization problem. *In The 2011 International Joint Conference on Neural Networks* (pp. 580-587). IEEE.

[53] Boccato, L., Lopes, A., Attux, R. and Von Zuben, F.J., 2012. An extended echo state network using Volterra filtering and principal component analysis. *Neural Networks*, 32, pp.292-302.

[53] Kuo, S.M. and Wu, H.T., 2005. Nonlinear adaptive bilinear filters for active noise control systems. *IEEE Transactions on Circuits and Systems I: Regular Papers*, *52*(3), pp.617-624.

[54] Luo, L. and Sun, J., 2018. A novel bilinear functional link neural network filter for nonlinear active noise control. *Applied Soft Computing*, *68*, pp.636-650.

[55] Zhang, J. and Zhao, H., 2010. A novel adaptive bilinear filter based on pipelined architecture. *Digital Signal Processing*, *20*(1), pp.23-38.

[56] Dong, C., Ding, Y., Tan, L., Du, S. and Guo, X., 2020. Diagonal-structure adaptive bilinear filters for multichannel active noise control of nonlinear noise processes. *Mechanical Systems and Signal Processing*, *143*, p.106703.

[57] Le, D.C., Li, D. and Zhang, J., 2019. M-max partial update leaky bilinear filter-error least mean square algorithm for nonlinear active noise control. *Applied Acoustics*, *156*, pp.158-165.

[58] Moore II, E., Clements, M.A., Peifer, J.W. and Weisser, L., 2007. Critical analysis of the impact of glottal features in the classification of clinical depression in speech. *IEEE transactions on biomedical engineering*, *55*(1), pp.96-107.

[59] Bartlett, P.L., 1998. The sample complexity of pattern classification with neural networks: the size of the weights is more important than the size of the network. *IEEE transactions on Information Theory*, *44*(2), pp.525-536.

[60] Yang, J. and Yang, J.Y., 2002. From image vector to matrix: A straightforward image projection technique—IMPCA vs. PCA. *pattern Recognition*, *35*(9), pp.1997-1999.

[61] Burkhardt, F., Paeschke, A., Rolfes, M., Sendlmeier, W.F. and Weiss, B., 2005. A database of German emotional speech. In *Ninth European Conference on Speech Communication and Technology*.

[62] Haq, S. and Jackson, P.J., 2011. Multimodal emotion recognition. In *Machine audition: principles, algorithms and systems* (pp. 398-423). IGI Global.

[63] Busso, C., Bulut, M., Lee, C.C., Kazemzadeh, A., Mower, E., Kim, S., Chang, J.N., Lee, S. and Narayanan, S.S., 2008. IEMOCAP: Interactive emotional dyadic motion capture database. *Language resources and evaluation*, *42*(4), p.335.

[64] Chen, M., He, X., Yang, J. and Zhang, H., 2018. 3-D convolutional recurrent neural networks with attention model for speech emotion recognition. *IEEE Signal Processing Letters*, *25*(10), pp.1440-1444.

[65] Sidorov, M., Minker, W. and Semenkin, E.S., 2016. Speech-based emotion recognition and speaker identification: static vs. dynamic mode of speech representation. *Journal of the Siberian Federal University. The series "Mathematics and Physics* , *9*(4), pp.518-523.

[66] Aghajani, K. and Esmaili Paeen Afrakoti, I., 2020. Speech Emotion Recognition Using Scalogram Based Deep Structure. *International Journal of Engineering*, *33*(2), pp.285-292

[67] Li, D., Zhou, Y., Wang, Z. and Gao, D., 2021. Exploiting the potentialities of features for speech emotion recognition. *Information Sciences*, *548*, pp.328-343

[68] Yüncü, E., Hacihabiboglu, H. and Bozsahin, C., 2014, August. Automatic speech emotion recognition using auditory models with binary decision tree and svm. In *2014 22nd International Conference on Pattern Recognition* (pp. 773-778). IEEE.

[69] Sinith, M.S., Aswathi, E., Deepa, T.M., Shameema, C.P. and Rajan, S., 2015, December. Emotion recognition from audio signals using Support Vector Machine. In *2015 IEEE Recent Advances in Intelligent Computational Systems (RAICS)* (pp. 139-144). IEEE.

[70] Deb, S. and Dandapat, S., 2016, June. Emotion classification using residual sinusoidal peak amplitude. In *2016 International Conference on Signal Processing and Communications (SPCOM)* (pp. 1-5). IEEE.

[71] Deb, S. and Dandapat, S., (a) 2017, March. Exploration of phase information for speech emotion classification. In *2017 Twenty-third National Conference on Communications (NCC)* (pp. 1-5). IEEE.

[72] Deb, S. and Dandapat, S., (b) 2017. Emotion classification using segmentation of vowel-like and non-vowel-like regions. *IEEE Transactions on Affective Computing*.

[73] Deb, S. and Dandapat, S., 2018. Multiscale amplitude feature and significance of enhanced vocal tract information for emotion classification. *IEEE transactions on cybernetics*, *49*(3), pp.802-815

[74] Tao, H., Liang, R., Zha, C., Zhang, X. and Zhao, L., 2016. Spectral features based on local Hu moments of Gabor spectrograms for speech emotion recognition. *IEICE TRANSACTIONS on Information and Systems*, *99*(8), pp.2186-2189.

[75] Shirani, A. and Nilchi, A.R.N., 2016. Speech Emotion Recognition based on SVM as Both Feature Selector and Classifier. *International Journal of Image, Graphics & Signal Processing*, *8*(4).

[76] Rintala, J., 2020. Speech Emotion Recognition from Raw Audio using Deep Learning

[77] Hassan, A. and Damper, R.I., 2012. Classification of emotional speech using 3DEC hierarchical classifier. *Speech Communication*, *54*(7), pp.903-916.

[78] Daneshfar, F., Kabudian, S.J. and Neekabadi, A., 2020. Speech emotion recognition using hybrid spectral-prosodic features of speech signal/glottal waveform, metaheuristic-based dimensionality reduction, and Gaussian elliptical basis function network classifier. *Applied Acoustics*, *166*, p.107360. https://doi.org/10.1016/j.apacoust.2020.107360

[79] Daneshfar, F. (a) and Kabudian, S.J., 2020. Speech emotion recognition using discriminative dimension reduction by employing a modified quantum-behaved particle swarm optimization algorithm. *Multimedia Tools and Applications*, *79*(1), pp.1261-1289. https://doi.org/10.1007/s11042-019-08222-8

[80] Zao, L., Cavalcante, D. and Coelho, R., 2014. Time-frequency feature and AMS-GMM mask for acoustic emotion classification. *IEEE signal processing letters*, *21*(5), pp.620-624.

[81] Bhargava, M. and Polzehl, T., 2013. Improving automatic emotion recognition from speech using rhythm and temporal feature. *arXiv preprint arXiv:1303.1761*.

[82] Badshah, A.M., Rahim, N., Ullah, N., Ahmad, J., Muhammad, K., Lee, M.Y., Kwon, S. and Baik, S.W., 2019. Deep features-based speech emotion recognition for smart affective services. *Multimedia Tools and Applications*, *78*(5), pp.5571-5589.

[83] Zhang, S., Zhao, X. and Lei, B., 2013. Speech emotion recognition using an enhanced kernel isomap for human-robot interaction. *International Journal of Advanced Robotic Systems*, *10*(2), p.114.

[84] Wu, S., Falk, T.H. and Chan, W.Y., 2011. Automatic speech emotion recognition using modulation spectral features. *Speech communication*, *53*(5), pp.768-785.

[85] Sun, Y., Wen, G. and Wang, J., 2015. Weighted spectral features based on local Hu moments for speech emotion recognition. *Biomedical signal processing and control*, *18*, pp.80-90.

[86] Xu, X., Deng, J., Zheng, W., Zhao, L. and Schuller, B., 2015. Dimensionality reduction for speech emotion features by multiscale kernels. In *Sixteenth Annual Conference of the International Speech Communication Association*.

[87] Man-Wai, Mak, 2016. Feature Selection and Nuisance Attribute Projection for Speech Emotion Recognition, *Technical Report and Lecture Note Series, Department of Electronic and Information Engineering, The Hong Kong Polytechnic University, Dec. 2016*.

[88] Stuhlsatz, A., Meyer, C., Eyben, F., Zielke, T., Meier, G. and Schuller, B., 2011, May. Deep neural networks for acoustic emotion recognition: Raising the benchmarks. In *2011 IEEE international conference on acoustics, speech and signal processing (ICASSP)* (pp. 5688-5691). IEEE.

[89] Lotfidereshgi, R. and Gournay, P., 2017, March. Biologically inspired speech emotion recognition. In *2017 IEEE International Conference on Acoustics, Speech and Signal Processing (ICASSP)* (pp. 5135-5139). IEEE.

[90] Sun, L., Fu, S. and Wang, F., 2019. Decision tree SVM model with Fisher feature selection for speech emotion recognition. *EURASIP Journal on Audio, Speech, and Music Processing*, *2019*(1), p.2.

[91] Sun, T.W. and Wu, A.Y.A., 2019, March. Sparse Autoencoder with Attention Mechanism for Speech Emotion Recognition. In *2019 IEEE International Conference on Artificial Intelligence Circuits and Systems (AICAS)* (pp. 146-149). IEEE.

[92] Sun, T.W., 2020. End-to-End speech emotion recognition with gender information. *IEEE Access*, *8*, pp.152423-152438

[93] Tzinis, E., Paraskevopoulos, G., Baziotis, C. and Potamianos, A., 2018. Integrating Recurrence Dynamics for Speech Emotion Recognition. *Proc. Interspeech 2018*, pp.927-931.

[94] Tzinis, E. and Potamianos, A., 2017, October. Segment-based speech emotion recognition using recurrent neural networks. In *2017 Seventh International Conference on Affective Computing and Intelligent Interaction (ACII)* (pp. 190-195). IEEE.

[95] Kalinli, O., 2016. Analysis of Multi-Lingual Emotion Recognition Using Auditory Attention Features. In *INTERSPEECH* (pp. 3613-3617).

[96] Zhang, S., Zhang, S., Huang, T. and Gao, W., 2017. Speech emotion recognition using deep convolutional neural network and discriminant temporal pyramid matching. *IEEE Transactions on Multimedia*, *20*(6), pp.1576-1590.

[97] Yi, L. and Mak, M.W., 2020. Improving Speech Emotion Recognition With Adversarial Data Augmentation Network. *IEEE Transactions on Neural Networks and Learning Systems*.

[98] Tawari, A. and Trivedi, M.M., 2010. Speech emotion analysis: Exploring the role of context. *IEEE Transactions on multimedia*, *12*(6), pp.502-509.

[99] Özseven, T., 2018. Investigation of the effect of spectrogram images and different texture analysis methods on speech emotion recognition. *Applied Acoustics*, *142*, pp.70-77.

[100] Özseven, T., 2019. A novel feature selection method for speech emotion recognition. *Applied Acoustics*, *146*, pp.320-326.

[101] Meng, H., Yan, T., Yuan, F. and Wei, H., 2019. Speech emotion recognition from 3D log-mel spectrograms with deep learning network. *IEEE Access*, *7*, pp.125868-125881.

[102] Papakostas, M., Spyrou, E., Giannakopoulos, T., Siantikos, G., Sgouropoulos, D., Mylonas, P. and Makedon, F., 2017. Deep visual attributes vs. hand-crafted audio features on multidomain speech emotion recognition. *Computation*, *5*(2), p.26.

[103] Liu, Z.T., Xie, Q., Wu, M., Cao, W.H., Mei, Y. and Mao, J.W., 2018. Speech emotion recognition based on an improved brain emotion learning model. *Neurocomputing*, *309*, pp.145-156.

[104] Noroozi, F., Marjanovic, M., Njegus, A., Escalera, S. and Anbarjafari, G., 2017. Audio-visual emotion recognition in video clips. *IEEE Transactions on Affective Computing*, *10*(1), pp.60-75.

[105] Noroozi, F., Sapiński, T., Kamińska, D. and Anbarjafari, G., 2017. Vocal-based emotion recognition using random forests and decision tree. *International Journal of Speech Technology*, *20*(2), pp.239-246.

[106] Vasquez-Correa, J.C., Arias-Vergara, T., Orozco-Arroyave, J.R., Vargas-Bonilla, J.F. and Noeth, E., 2016, October. Wavelet-based time-frequency representations for automatic recognition of emotions from speech. In *Speech Communication; 12. ITG Symposium* (pp. 1-5). VDE.

[107] Nguyen, D., Sridharan, S., Nguyen, D.T., Denman, S., Tran, S.N., Zeng, R. and Fookes, C., 2020. Joint Deep Cross-Domain Transfer Learning for Emotion Recognition. *arXiv preprint arXiv:2003.11136*

[108] Jiang, P., Fu, H., Tao, H., Lei, P. and Zhao, L., 2019. Parallelized Convolutional Recurrent Neural Network With Spectral Features for Speech Emotion Recognition. *IEEE Access*, *7*, pp.90368-90377.

[109] Farooq, M., Hussain, F., Baloch, N.K., Raja, F.R., Yu, H. and Zikria, Y.B., 2020. Impact of Feature Selection Algorithm on Speech Emotion Recognition Using Deep Convolutional Neural Network. *Sensors*, *20*(21), p.6008

[110] Latif, S., Rana, R., Khalifa, S., Jurdak, R. and Epps, J., 2019. Direct Modelling of Speech Emotion from Raw Speech. *Proc. Interspeech 2019*, pp.3920-3924.

[111] Zong, Z., Li, H. and Wang, Q., 2018. Multi-Channel Auto-Encoder for Speech Emotion Recognition. *arXiv preprint arXiv:1810.10662*.

[112] Kwon, S., 2020. A CNN-Assisted Enhanced Audio Signal Processing for Speech Emotion Recognition. *Sensors*, *20*(1), p.183.

[113] Ghosh, S., Laksana, E., Morency, L.P. and Scherer, S., 2016, September. Representation Learning for Speech Emotion Recognition. In *Interspeech* (pp. 3603-3607).

[114] Xie, Y., Liang, R., Liang, Z. and Zhao, L., 2019. Attention-Based Dense LSTM for Speech Emotion Recognition. *IEICE TRANSACTIONS on Information and Systems*, *102*(7), pp.1426-1429.

[115] Li, P., Song, Y., McLoughlin, I., Guo, W. and Dai, L., 2018. An Attention Pooling Based Representation Learning Method for Speech Emotion Recognition. *Proc. Interspeech 2018*, pp.3087-3091

[116] Li, Y., Chao, L., Liu, Y., Bao, W. and Tao, J., 2015, September. From simulated speech to natural speech, what are the robust features for emotion recognition?. In *2015 International Conference on Affective Computing and Intelligent Interaction (ACII)* (pp. 368-373). IEEE.

[117] Zhao, H., Xiao, Y., Han, J. and Zhang, Z., 2019, May. Compact Convolutional Recurrent Neural Networks via Binarization for Speech Emotion Recognition. In *ICASSP 2019-2019 IEEE International Conference on Acoustics, Speech and Signal Processing (ICASSP)* (pp. 6690-6694). IEEE.

[118] Zhao, Z., Zheng, Y., Zhang, Z., Wang, H., Zhao, Y. and Li, C., 2018. Exploring spatio-temporal representations by integrating attention-based bidirectional-LSTM-RNNs and FCNs for speech emotion recognition.

[119] Huang, J., Li, Y., Tao, J. and Lian, Z., 2018, September. Speech Emotion Recognition from Variable-Length Inputs with Triplet Loss Function. In *Interspeech* (pp. 3673-3677).

[120] Mao, S., Tao, D., Zhang, G., Ching, P.C. and Lee, T., 2019, May. Revisiting hidden Markov models for speech emotion recognition. In *ICASSP 2019-2019 IEEE International Conference on Acoustics, Speech and Signal Processing (ICASSP)* (pp. 6715-6719). IEEE.

[121] Han, W., Ruan, H., Chen, X., Wang, Z., Li, H. and Schuller, B.W., 2018. Towards Temporal Modelling of Categorical Speech Emotion Recognition. In *Interspeech* (pp. 932-936).

[122] Issa, D., Demirci, M.F. and Yazici, A., 2020. Speech emotion recognition with deep convolutional neural networks. *Biomedical Signal Processing and Control*, *59*, p.101894

[123] Etienne, C., Fidanza, G., Petrovskii, A., Devillers, L. and Schmauch, B., 2018. CNN+ LSTM Architecture for Speech Emotion Recognition with Data Augmentation. In *Proc. Workshop on Speech, Music and Mind 2018* (pp. 21-25).

[124] Fayek, H.M., Lech, M. and Cavedon, L., 2017. Evaluating deep learning architectures for Speech Emotion Recognition. *Neural Networks*, *92*, pp.60-68.

[125] Song, P., Zheng, W., Yu, Y. and Ou, S., 2020. Speech Emotion Recognition Based on Robust Discriminative Sparse Regression. *IEEE Transactions on Cognitive and Developmental Systems*.

[126] Er, M.B., 2020. A Novel Approach for Classification of Speech Emotions Based on Deep and Acoustic Features. *IEEE Access*, *8*, pp.221640-221653.

[127] Sajjad, M. and Kwon, S., 2020. Clustering-Based Speech Emotion Recognition by Incorporating Learned Features and Deep BiLSTM. *IEEE Access*, *8*, pp.79861-79875.

[128] Kwon, S. 2021, Att-Net: Enhanced emotion recognition system using lightweight self-attention module. *Applied Soft Computing*, p.107101.

[129] Bakhshi, A., Chalup, S., Harimi, A. and Mirhassani, S.M., 2020. Recognition of emotion from speech using evolutionary cepstral coefficients. *Multimedia Tools and Applications*, *79*(47), pp.35739-35759..